\documentclass[epj,final]{svjour} 
\usepackage{graphicx,epsf}
\title{Numerical study of homogeneous dynamo based 
on experimental von K\'arm\'an type flows}

\author{L. Mari\'e\inst{1} \and J. Burguete\inst{1,2} \and 
	F. Daviaud\inst{1} \and J. L\'eorat\inst{3,4} }

\institute{Service de Physique l'\'Etat Condens\'e,
		CEA/Saclay, F-91191 Gif-sur-Yvette, France \and
	Departamento de F\'{\i}sica y Matem\'atica Aplicada,
		Universidad de Navarra, E-31080 Pamplona, Spain \and
	Observatoire de Paris-Meudon, F-92195 Meudon, France \and
	ASCI, Universit\'e Paris Sud, F-91403 Orsay, France}

\date{\today}

\abstract{
A numerical study of the magnetic induction equation 
has been performed on von K\'arm\'an type flows. These flows are generated
by two co-axial counter-rotating propellers in cylindrical containers.
Such devices are currently used in the  von K\'arm\'an sodium (VKS)
experiment designed to study dynamo action in an unconstrained flow.
The mean velocity fields have been measured for different configurations
and are introduced in a periodic cylindrical kinematic dynamo code.
Depending on the driving configuration, on the poloidal to toroidal
flow ratio and on the conductivity of boundaries, some flows are observed to 
sustain growing  magnetic fields for magnetic Reynolds numbers accessible
to a sodium experiment. The response of the flow to an external magnetic field
has also been studied: The results are in excellent agreement with experimental 
results in the single propeller case but can differ in the two propellers case.}


\begin{document}

\maketitle

\section{Introduction}

Dynamo action~\cite{Mof78,Mor90}, which converts kinetic energy into magnetic energy in
astronomical objects, is the manifestation of the coupling between kinetic
and magnetic excitations in a conducting fluid and, as such, could be
considered as ordinary a physical phenomenon as is thermal 
convection. This is in fact not the case, when one considers
 the modest knowledge acquired from all the approaches carried out up to now
 from theory, numerical computation or experiments. Since the recent results
obtained by the Riga~\cite{Gai00,Gai01}  and Karlsruhe~\cite{Sti00,Sti01}
experiments, the occurence of dynamo action is no longer questionable,
but the nonlinear regimes are very poorly known: most basic problems set
for example by geomagnetism or heliomagnetism remain without satisfying
answers.

As for hydrodynamic turbulence, the experimental approach could represent
an efficient tool to study the nonlinear effects in MHD flows. We will here
concentrate on an analysis of a particular type of flow - namely von K\'arm\'an
flow -  without trying to give a complete account of this field of
activity. These flows are used in the von K\'arm\'an sodium (VKS) 
experiment~\cite{Mar01,Bou02} that is devoted to study the approach towards a 
self-generating dynamo in an unconstrained flow. No self-excitation has been
reported yet and the ability of these flows to generate a magnetic field remains
an open question. Some numerical examples~\cite{Dud89} 
have shown that dynamo action is present in flows for magnetic Reynolds numbers
$R_{\rm m} = \mu_0 \sigma U  L \ge R_{\rm m}^c \simeq 100$ where
$U $ and $L $ are respectively the maximal speed of 
the flow and the typical size of the conducting volume,
$\mu_0$ is the magnetic permeability and $\sigma$ the electrical conductivity. 
 Using the best available fluid
conductor, liquid sodium at about $150^{\circ}$ C, the condition $R_{\rm m} = 100$
 implies that $U L = 10 m^2/s$,
 which represents the main technical challenge
to be achieved by any experimental fluid dynamo. In the natural dynamos,
such as the Earth, large magnetic Reynolds numbers are achieved with scales
above 1000 km and small velocities, while in an experimental device, cost
constraints ask for sizes not too far from one meter and thus a relatively
high speed is needed.

Note also that using liquid sodium as a conducting fluid, the kinetic
Reynolds number of the flow is  $Re = U L/\nu = R_{\rm m}/P_{\rm m}$, where
$P_{\rm m}= \mu_0 \sigma \nu$ is the magnetic Prandtl number and $\nu$ the 
kinematic viscosity. As $P_{\rm m} \sim 10^{-5}$ for liquid metals, 
 $Re \simeq 10^5 \times R_m \simeq 10^7$,
 which shows that the flow is in a regime of fully developped turbulence.
 Moreover, as is well known, the threshold $R_{\rm m}^c$ for dynamo action
depends strongly on the characteristics of the flow. Three different methods 
may then be used to select an experimental configuration:\\
(i) reproduce flows already known to lead to dynamo action, based on
previous theoretical or numerical knowledge. Very recently, this approach
has been successfully used in constrained flows in recent experiments
in Riga~\cite{Gai01}, based on the Ponomarenko model~\cite{Pon72}, 
 and Karlsruhe~\cite{Sti01}, based on the G. O. Roberts 
model~\cite{Rob72}. However, imposing the desired flow topology
may require internal walls, which may have an influence on the non-linear
regime also. Despite different attempts~\cite{Pef00,Pef01,Mar01,Bou02}, 
dynamo action in an unconstrained turbulent flow has not yet been observed.\\
(ii) try various forcing mechanisms and geometries using directly a
conducting flow prototype, and determine the corresponding critical
magnetic Reynolds numbers (e.g. by measuring the decay times
of an external magnetic field or the response to an external magnetic field).
This approach  has been used in sodium by Gans~\cite{Gan70}, 
by Odier et al.~\cite{Odi99} in gallium and  by Peffley et al.~\cite{Pef00,Pef01} 
in the sodium Maryland experiment. 
The main drawback is  the lack of knowledge of the velocity field: if the
selected configuration leads to a threshold $R_{\rm m}^c$ being too high to be
feasible, there is no guide besides trials and errors to make it smaller.\\
(iii) study various forcing mechanisms and geometries in water models to
measure the mean velocity fields which are then introduced in the numerical
computation of a kinematic dynamo problem. This is the way chosen in particular
by the Madison~\cite{For01}, the Perm~\cite{Fri01} and the VKS 
experiments~\cite{Mar01,Bou02}.

In this work, a numerical study of the induction equation is performed
on von K\'arm\'an type flows similar to those used in the VKS experiment.
These flows are generated by counter-rotating disks in a cylindrical geometry
and have been extensively studied in the past~\cite{Pic58,Wel58,Zan87,Fau93,Pin94,Mor97}.
They are supposed to be good candidates to the realization of an experimental
homogeneous fluid dynamo. In particular,
kinematic dynamo simulations in a sphere~\cite{Dud89} and direct numerical 
simulations of the Taylor-Green geometry~\cite{Nor97} have shown that similar flows
lead to self-excitation for accessible $R_{\rm m}^c$. 
No dynamo action has been presently observed in these unconstrained geometries, but
two types of MHD measurements have been performed: in a sphere filled with sodium,
Peffley et al.~\cite{Pef00} have used pulse-decay rates to obtain an estimation of
$R_{\rm m}^c$; the VKS experiment~\cite{Bou02} has studied the response of the
flow to an external field and exhibited large magnetic induction effects.
However, the dependence of the threshold on the different parameters of the problems
remains unknown. In this paper, the mean velocity fields are measured in a water model 
experiment for various configurations and introduced in an axially periodic cylindrical 
kinematic dynamo code. The dependence of the threshold on the main characteristics 
of the flow and on the boundary conditions is then studied.

The paper is organized as follows. The setup and the velocity fields of
the  model experiment are  presented in Section 2 and the numerical approach 
in Section 3. The determination of threshold, the description of the self-excited 
magnetic field and the response of the system to an external magnetic field are
then reported in Section 4. The results are finally discussed and compared to available
results in Section 5.


\section{Water model experiment}

\subsection{Experimental setup}

\begin{figure}[htbp]
\centering
\includegraphics[width=0.5 \textwidth,clip=true]{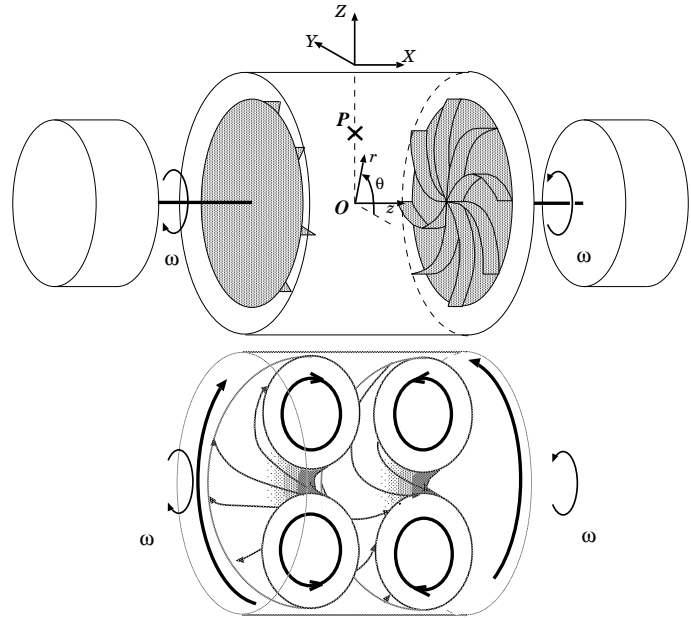}
\caption{Top: experimental setup. The flow is produced in a cylindrical container
by two counter-rotating propellers driven by two independent motors. 
(r, $\theta, z$) are the usual cylindrical coordinates. 
P represents the point where the magnetic field probe is located inside the 
VKS apparatus. (X, Y, Z) are the cartesian coordinates corresponding to the probe
measurement axes. Bottom: schematic drawing of the von K\'arm\'an flow showing 
the mean toroidal and poloidal flows.}
\label{Setup}
\end{figure}

The experimental setup shown in figure~\ref{Setup} is a half-scale model
of the sodium VKS experiment~\cite{Bou02}. It consists of a 
cylindrical container of internal radius $R_c=10$ cm  and 
height $H_c=30$ cm, filled with water. The system is driven by two counter-rotating
co-axial  disks  of radius $R_d$, with their inner faces a distance $H_d =18$ cm apart.
Different disks have been used: smooth or rough disks, disks with
straight or curved blades with different heights.  In the following, we report
results concerning two different propellers:
Propeller TM28 (resp. TM60) has a 180 mm diameter (resp. 185 mm), 
8 blades (resp. 16) of 2 cm height. The blades of TM60 have a curvature slightly 
larger than those of TM28.
Both propellers rotate as shown in figure~\ref{Setup}.
The propellers presently used in the VKS experiment are of
the TM60 type.
Baffles of different width and length can be introduced on the internal
wall of the cylinder in order to change the ratio of axial to azimuthal components
of the velocity field. 

Two 2 kW  motors are used to drive the  disks 
in opposite directions at an adjustable frequency $f$ in the range 0-25 Hz.
The measurements reported here have been performed in the range 0-10 Hz.
A kinematic Reynolds number based on the driving is defined as: 
$ Re = 2\pi f  R_c^2 /\nu $, where $\nu$ is the kinematic viscosity of water.

Global measurements such as torque or power and pressure measurements have been
performed in order to characterize the turbulence. The results will be reported 
in details elsewhere and show results typical of fully turbulent flows.
In the following, we focus on the velocity measurements which are used in the 
numerical study.  

\subsection{Velocity fields}\label{velsection}

\subsubsection{Measurement of the velocity field}

\begin{figure}[htbp]
\centering
\includegraphics[width=0.4 \textwidth,clip=true]{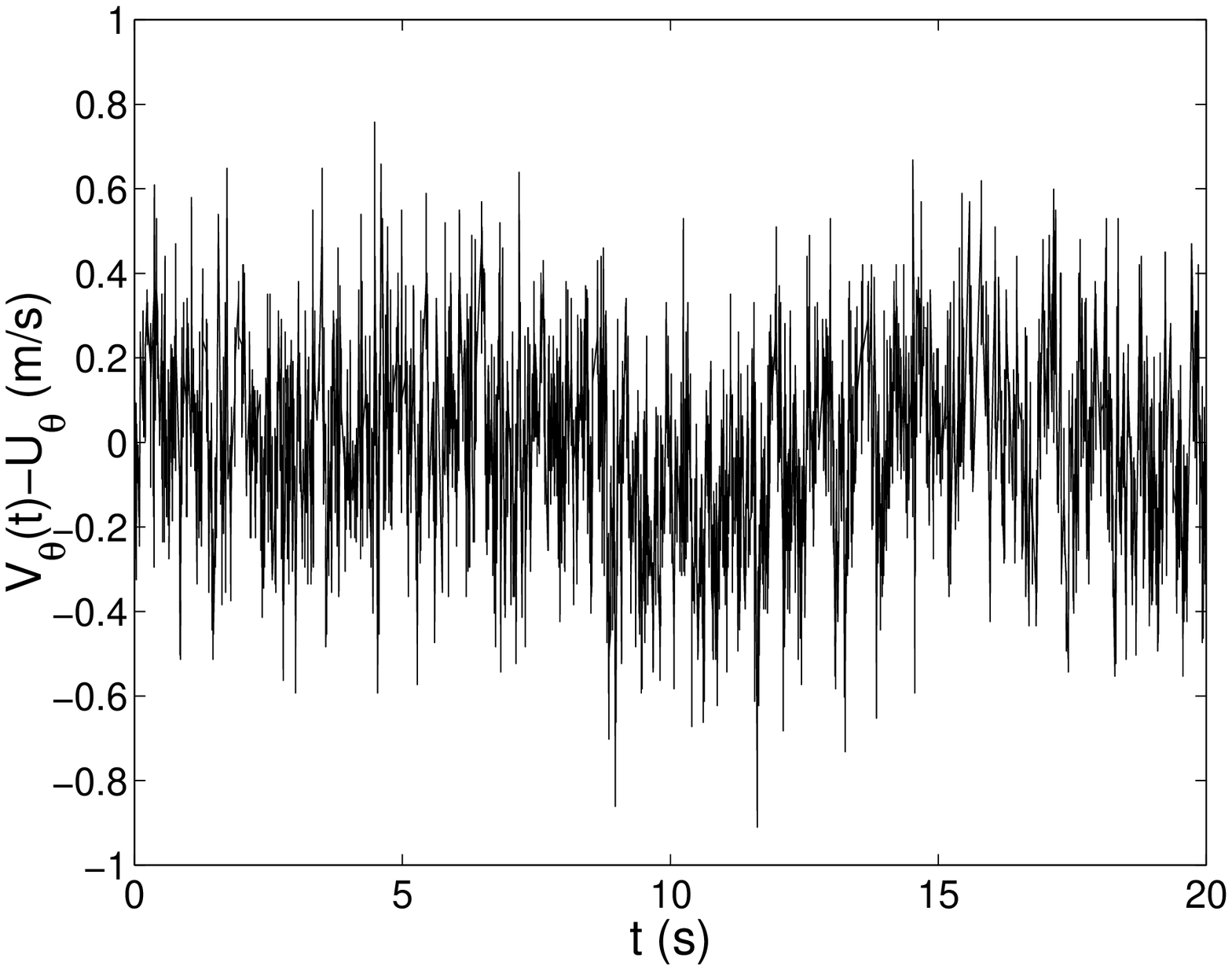}
\includegraphics[width=0.4 \textwidth,clip=true]{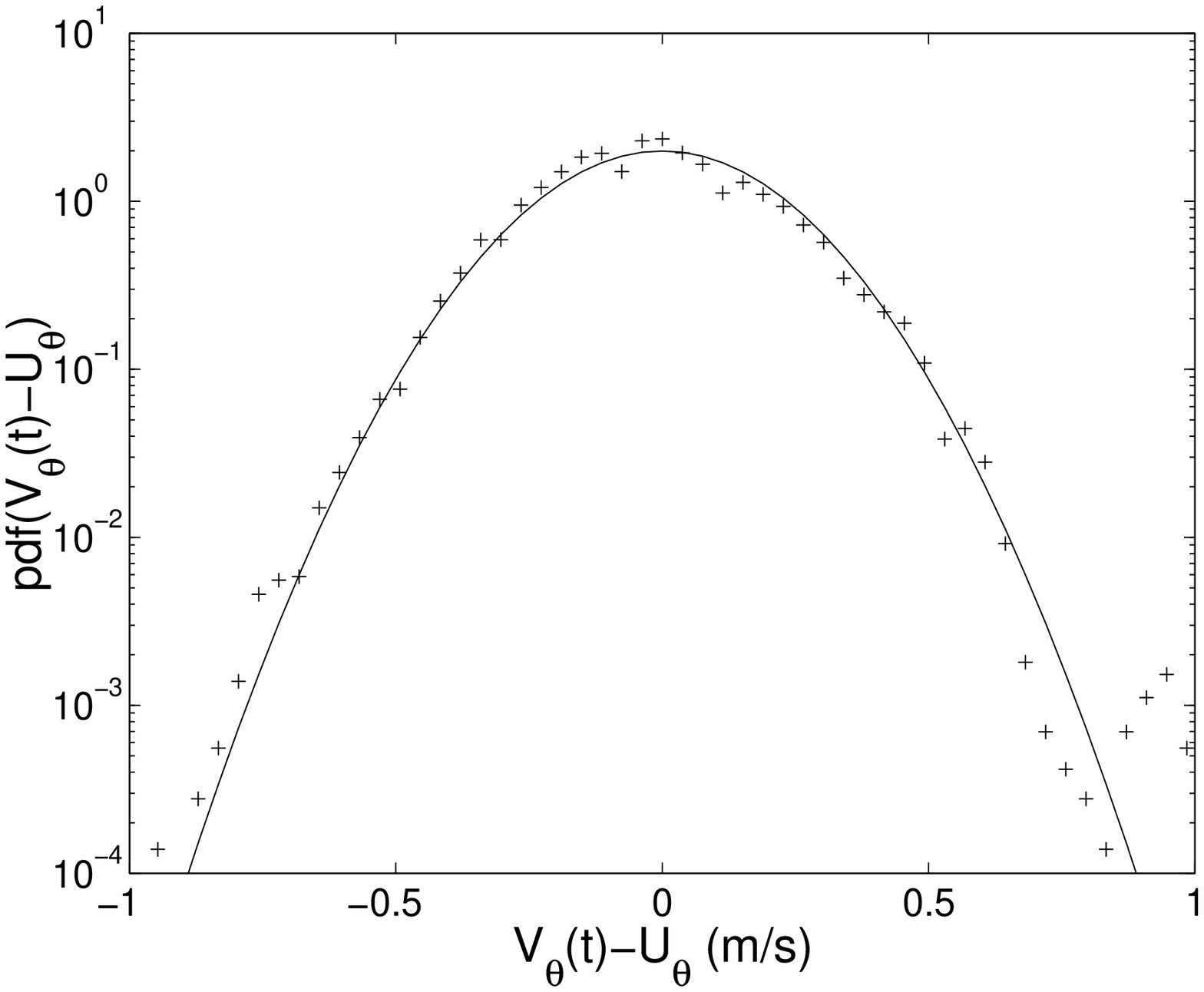}

\caption{(a) Temporal evolution of the fluctuating part of the
azimuthal velocity of TM60 propeller
recorded using Laser Doppler Velocimetry at $r=60$ mm, $z=65$ mm for $f=2$ Hz.
The value of the mean azimuthal velocity is $U_{\theta} = 0.42 \ {\rm ms}^{-1}$.
(b) Probability density function for the same signal: data ($+$) and gaussian 
fit (solid line).}
\label{TurbSpec}
\end{figure}

Velocity fields have been obtained via laser Doppler velocimetry 
(LDV) using a DANTEC system. The axial and azimuthal velocities 
$V_z$ and $V_\theta $ have been measured as functions of $r$, $z$ and time.
At the rotation rates used in the experiment, the flow is highly
turbulent: $Re \simeq 6. 10^5$ for $f= 10$ Hz.
Figure~\ref{TurbSpec} shows the temporal evolution of the local velocity
at a given point: velocity fluctuations appear to be of the order of magnitude
of the mean velocity, and have a quasi-gaussian distribution. In fact, the amplitude
of the fluctuations as well as the shape of their probability density function strongly 
depend on the position inside the flow. The amplitude is particularly
high in the plane between the two recirculation cells, where the strong toroidal shear
maintains a vigorous mixing layer.
More quantitatively, the turbulence intensity defined as 
$ K_V= V_{rms} /  U_{rim}$,
where $V_{rms}$ is the order of magnitude of the standard deviation of the velocity
throughout the flow and $U_{rim}$ is the rim velocity of the disks, is roughly 40 \%.

As a consequence, the velocity can be seen as a mean flow plus a turbulent part:
$ \vec V = \vec U + \vec u $. In the following, $\vec U$ denotes a 
time averaged flow,  and $\vec V$  a velocity field that depends on time. 
 
The LDV facility gives only the axial and azimuthal instantaneous velocity,
whereas the induction equation asks for the three time-averaged components. 
The missing mean radial velocity is derived as follows.
Note that, although $\vec U$ is not a solution of the Navier Stokes equations, 
it is a solenoidal vector field, and it can  always be decomposed into 
toroidal and poloidal components:
$$ \vec U (r, \theta, z) = \nabla \times (T  \hat z) + \nabla \times \nabla \times (P  \hat z) 
= \vec U_{ tor } + \vec U_{pol} $$
where $ \hat z$ is the unitary vector in the axial direction.
Assuming now that $ \vec U$ is axisymmetric (it is not strictly the case when 
four axial baffles are used), its toroidal component reduces to the azimuthal 
velocity, while its poloidal component is found to lie in the meridian planes
\begin{eqnarray*}
\vec U_{tor}(r,z) & = & U_\theta (r,z)\ \hat \theta \\
\vec U_{pol}(r,z) & = & \nabla \times (\partial_r P \ \hat \theta) =
                      \hat \theta \times (1/r\, \nabla \psi (r,z)) \\
\vec U_{pol}(r,z) & = & U_r (r,z)\ \hat r + U_z (r,z)\ \hat z 
\end{eqnarray*}
where $\hat r$ and $\hat \theta $ are the unitary vectors in the radial and 
azimuthal directions. The poloidal scalar P is thus replaced by the flux 
function $\psi (r,z)$, such that
\begin{eqnarray*}
U_r (r,z) & = &  1/r\, \partial_z \psi\\ 
U_z (r,z) & = & -1/r\, \partial_r \psi 
\end{eqnarray*}
so that finally $U_r$ can be derived from the experimental knowledge of the 
field $U_z (r,z) $.

The velocity field input in the kinematic dynamo code is obtained in the following way:
(i) The time averaged $U_\theta$ and $U_z$ are measured on a $11 \times 15$ grid in the plane $\theta=0$
of figure~\ref{Setup}, outside the region swept by the propeller blades.
(ii) In the region swept by the propeller blades, we interpolate lines $z=0,\ 16,\ 17$, making the 
hypothesis that $U_z$ depends linearly in $z$  and $U_\theta$ does not depend on $z$.
(iii) From the obtained $11 \times 18$ $U_z$ velocity field, we compute the flux function $\psi$ by
a radial integration between the axis and the container wall.
(iv) $\psi$ and $U_\theta$ are smoothed using a standard $ 3\times 3$ convolution  filter.
(v) $\psi$ and $U_\theta$ are then periodized in the z-direction in order to eliminate
Gibbs phenomenon in the code. Axial periodicity is obtained by
completing the measured flow by its symmetric with respect to the plane of any of the container tops
(as translation of length $L$ is obtained by the product of two reflections by parallel planes 
distant of $L/2$).
(vi) $\psi$ and $U_\theta$ are then interpolated linearly to the final $48(z) \times 51(r)$ 
simulation resolution.
(vii) $U_r$ is then derived from $\psi$.

Note that the periodization of the flow implies that the toroidal flow
can be decomposed on $n'$ odd modes, while the poloidal flow can be decomposed
on $n'$ even modes. The magnetic eigenmodes depend on these symmetry properties of the
simulated flow.


\subsubsection{Characterization of the velocity field} 

\begin{figure}[htbp]
\centering
\includegraphics[width=0.48 \textwidth,clip=true]{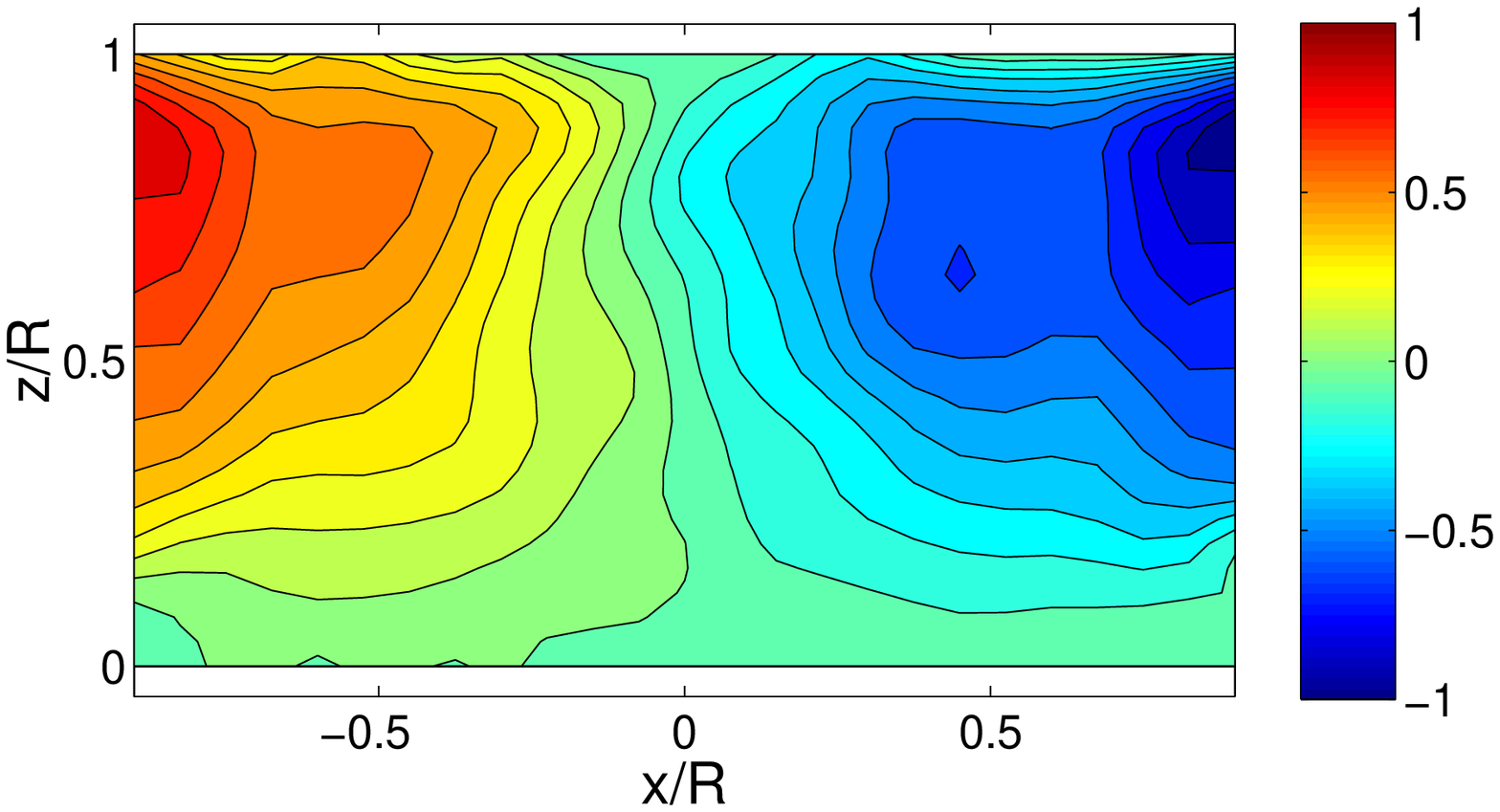}
\includegraphics[width=0.4 \textwidth,clip=true]{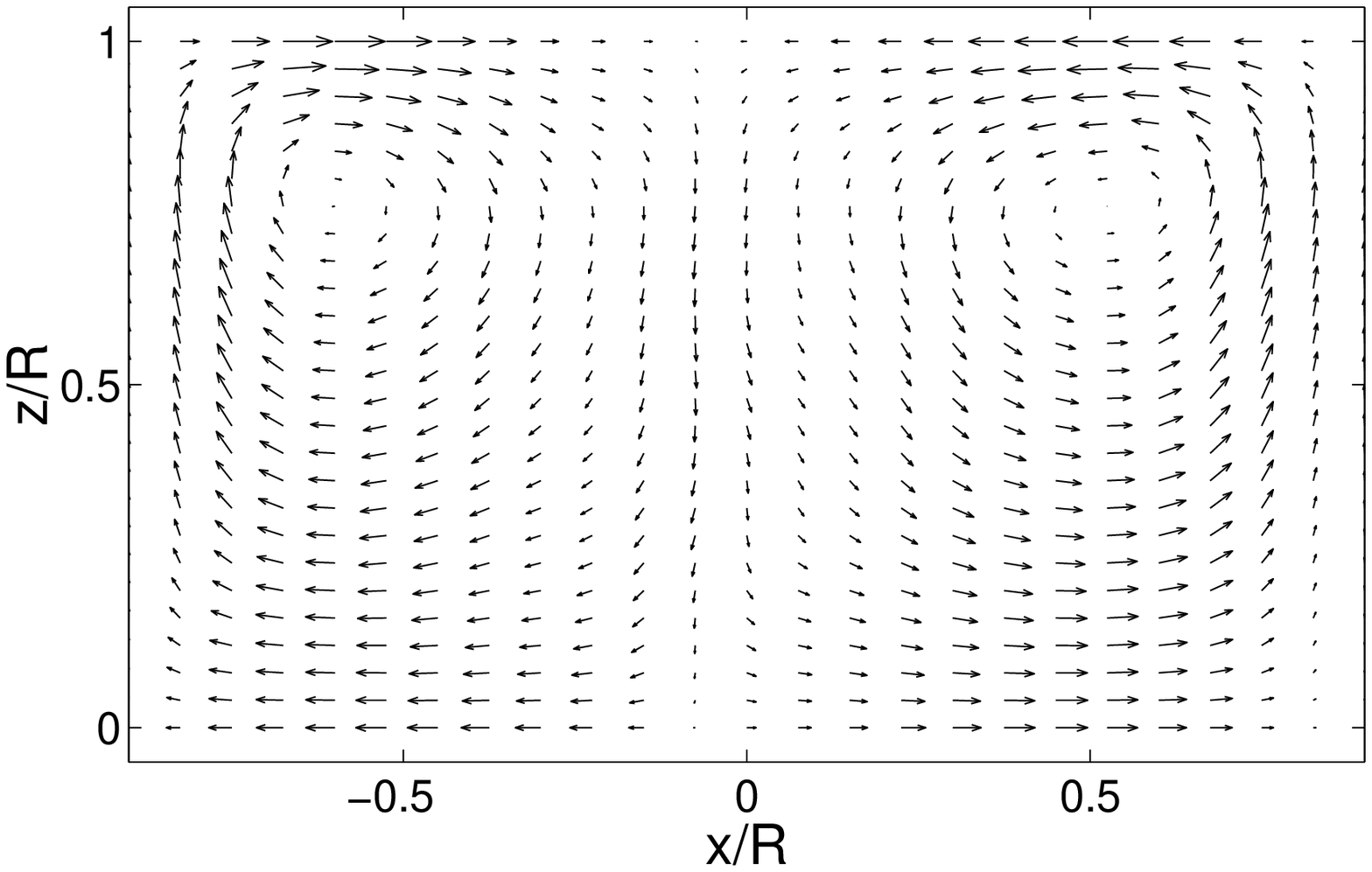}
\includegraphics[width=0.48 \textwidth,clip=true]{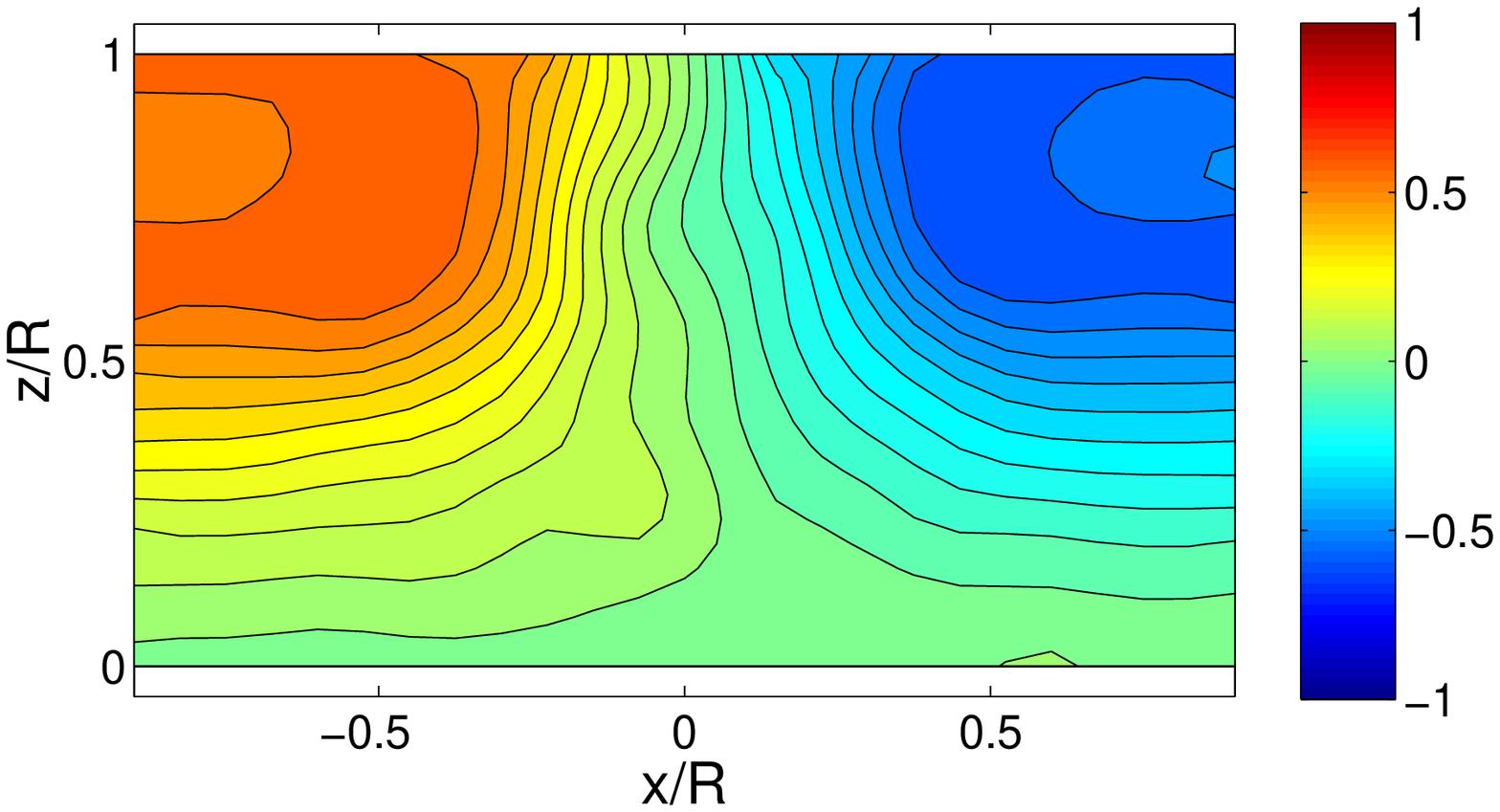}
\includegraphics[width=0.4 \textwidth,clip=true]{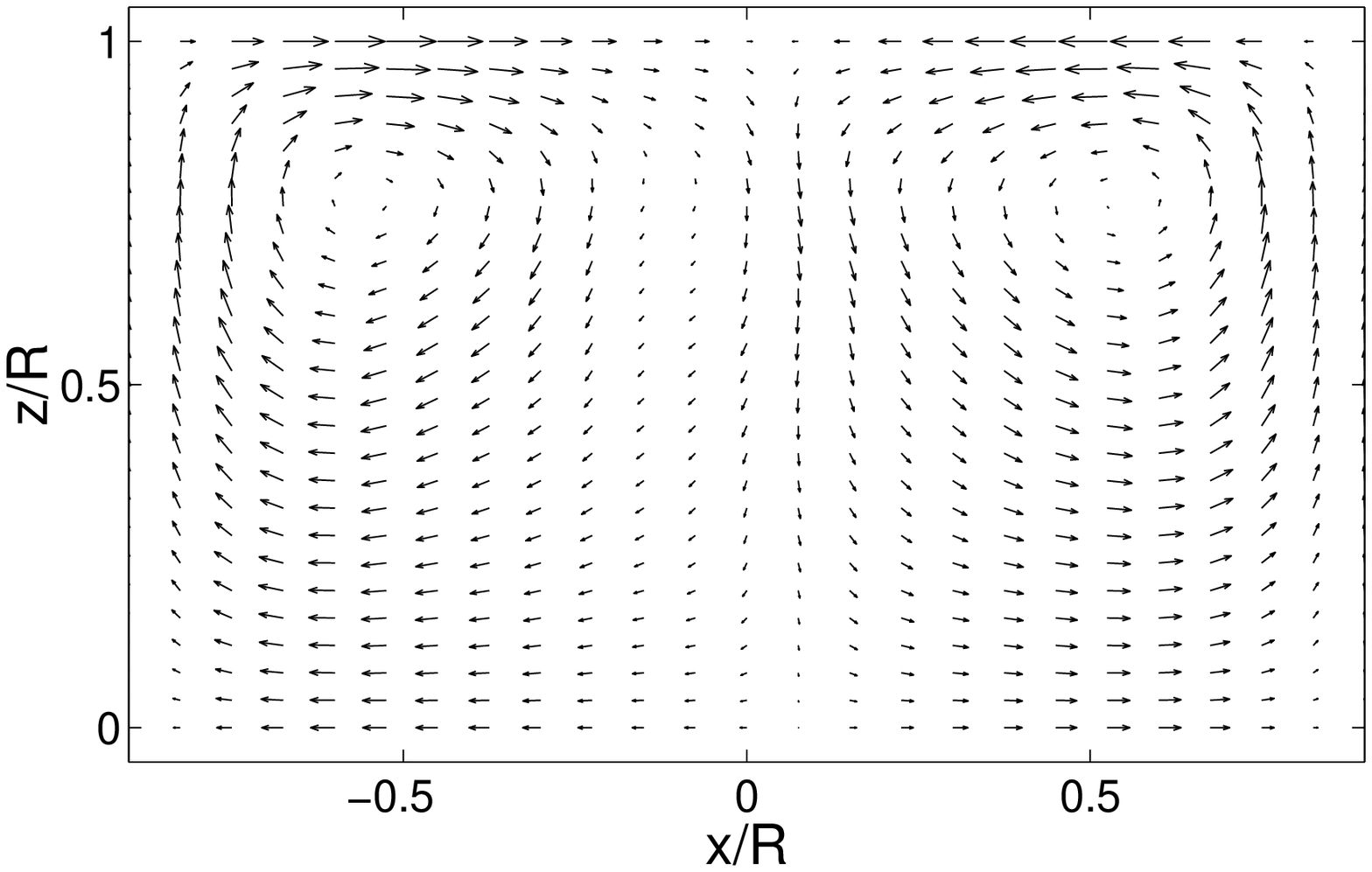}
\caption{Mean velocity field.  TM28 propeller velocity field:
(a) toroidal and (b) poloidal component in a meridional plane. 
TM60 propeller velocity field: (c) toroidal and (d) poloidal component.}
\label{MeanV}
\end{figure}

Different driving devices have been tested and each configuration 
gives a different velocity field. Figure~\ref{MeanV} shows the toroidal
and poloidal components of the mean velocity field for the 
propellers TM28  and TM60  without baffles. The two velocity fields appear to
be similar, but we will see later that TM28 and TM60 display quite different dynamo
properties.

In table~\ref{ComparativeVitesse}, the main characteristics of these 
velocity fields are summarized.  The spatial mean speed in the 
measured volume, including blades,  is defined as:
$$U_{mean} = {1\over V} \int_V | \vec U ( r, z ) | dV $$ 
and the  efficiency $ E_f$ of the propeller as: $ E_f = U_{\rm max}/ 2\pi R_c f$. 
$U_{max}$ represents the maximum value of the speed in the measured volume.
Typically, $U_{max} \leq  U_{\rm rim}$.

As can be seen in table~\ref{ComparativeVitesse}, the efficiency of propeller
TM28 is 25\% larger than that of propeller TM60. Note that the efficiency of a
straight-blade propeller is closer to unity.
As could be expected from the curvature difference, the poloidal-to-toroidal ratio
of propeller TM60 is larger than that of propeller TM28. This gain in
poloidal-to-toroidal ratio is however offset by the loss in efficiency, so that the
poloidal flow of TM60 is smaller than that of TM28.

The largest values of the velocity can be found close to the propeller rims.
This is true for both the poloidal and the toroidal velocities, 
regardless of the propeller used.

For a typical pair of propellers, we have performed velocity field measurements
at different rotation rates, 2.5, 5 and 7.5 Hz. At every location inside 
the flow, the measured velocity is proportional to the rotation rate.
This result could be expected from standard hydrodynamics arguments.
As a further consequence, changing the magnetic Reynolds number by a
simple scaling of the flow seems reasonable. This property will be
used in the numerical code.


\section{Numerical resolution of the induction equation}

\subsection{Scope of the numerical approach}
The numerical dynamo problem asks for the resolution of two 
coupled sets of equations, one for the velocity field and 
one for the induction equation. In the context of 
experimental dynamos using liquid sodium, at $R_{\rm m} = 100$,
the kinetic Reynolds number of the flow will reach 
$10^7$, which is far out of range of any direct numerical 
simulation. This is why scaled down water experiments are 
needed to measure the mean flow velocity field, to estimate 
the mechanical torque necessary to drive the flow as well 
as the dissipated power. Although the chaotic properties of 
the flow may play an essential role for the dynamo action in 
specific configurations, there is presently no way to 
determine the time dependent turbulent velocity field. 

An alternative to the fully nonlinear dynamo solution is the 
"kinematic" approach, where the flow is considered as given
in the evolution equation of the magnetic field. Starting 
with a faint seed field, this approximation is valid as long 
as the magnetic force field remains small. 
A better contact with experiments may be obtained if the 
time dependent solutions are obtained, instead of choosing to 
solve an eigenvalue problem. In this case, for flows at 
magnetic Reynolds numbers below the critical
value, one may study the magnetic response to external 
magnetic fields, possibly time-dependent, in order to get 
an experimental estimate of the critical $R_{\rm m}$. 
Finally, note that while the nonlinear simulations need a 
challenging  amount of numerical  ressources, the parameter 
space may be explored more rapidly and thoroughly using the 
kinematic solution. When following an optimization 
procedure, this is a meaningful practical point. A typical 
example is  given below by the variation of the thickness of 
an high conductivity blanket, which allows to reduce the 
critical $R_{\rm m}$. 
\subsection{The kinematic dynamo code }
As explained above, we consider that the velocity field is 
known in a cylindrical container of radius $R_c$. Moreover we assume that 
the conductivity $\sigma$ of the fluid inside the cylinder 
is uniform, the external medium is insulating and that the 
magnetic permeability is uniform in all space.

The induction equation is adimensionalized using the 
cylinder radius as length unit and the ohmic diffusion time $t_d = \mu_0 \sigma R_c^2$
as time unit. The induction equation reads then
$$ \partial_t \vec B = R_{\rm m} \nabla \times \left( \vec U 
\times \vec B \right) + \nabla ^2 \vec B.$$
Notice that the initial magnetic field must satisfy
$$ \nabla . \vec B = 0.$$
Boundary conditions are implemented as follows.
The magnetic field must be continuous at the cylinder 
boundary, from Maxwell equations. The external field is 
curl-free and derives from an harmonic scalar potential,
which is completely defined by, say, its gradient normal to 
the cylinder, i.e. the magnetic field orthogonal to the 
boundary. For a given conducting volume, the numerical
determination of the external harmonic potential from its 
gradient at the surface boundary may involve substantial
numerical resources, and this is in particular the case for 
a cylinder of finite length. We have chosen here to avoid 
this problem by solving the induction equation for axially 
periodic flows, where there is an analytical solution for 
the external potential, as is also the case for the 
spherical geometry. 

The magnetic field has thus the following representation,
$$
\vec B ( r,\varphi,z,t)=
	\sum_{n,m} \vec B^{n,m}(r,t) \exp\left[ i(m \theta + n z )\right], $$
where the z coordinate ($ 0 \leq z \leq 2 \pi $) has been 
scaled with the axial period and the integers n et m 
characterize the axial and azimuthal modes.
The spatial scheme is pseudospectral in the azimuthal and 
axial directions, and uses compact finite differences in the radial direction.

Let us summarize now the formal organisation of the temporal 
scheme. Suppose that at time $ t^i$ the internal magnetic field 
${\bf b}({\bf r}, t^i)$ is known: from the normal component of the surface 
field ${\bf b}_\perp ({\bf S}, t^i)$, where ${\bf S}$ is a point on the surface 
$r=1$, one may get the external potential 
for the next time step, ${\phi ({\bf r}, t^{i+1}) (r>1)} $
and the corresponding tangential magnetic field is then  obtained by differentiation
${\bf b}_{\parallel} ({\bf S}, t^{i+1}) = 
\vec \nabla_{\parallel} \phi({\bf S}, t^{i+1})$. These two components are the
two boundary conditions used for the integration of the dynamo equation within
the cylinder, which gives finally the internal field
${\bf b}({\bf r}, t^{i+1}), (r<1) $ at time step $i+1$.
The time scheme is second order Adams-Bashforth for the non linear 
terms, while the purely multiplicative parts of the diffusive terms are 
integrated exactly.
The integration variables are the three components of the 
magnetic field, and this allows to follow the growth of the 
divergence of the magnetic field, whose solenoidal property 
is not preserved under this algorithm (the discretized 
expression of div curl is not zero). To keep the divergence 
small, the numerical solution is projected, every 40 steps, 
say, on a divergenceless field. Other features of the code 
have been described elsewhere~\cite{Leo98}.

Comparisons with a finite cylinder 
dynamo code (performed by F. Stefani, RFZ, Dresden) have not 
shown significant differences for the critical $R_{\rm m}$ 
with respect to the simpler periodic solution having an identical 
aspect ratio, at least for a few flow configurations which have been tested.

\section{Results}

Dynamo action may take place if the energy generation by stretching
dominates the ohmic dissipation. This is measured in a loose way by
the magnitude of the magnetic Reynolds number $ R_{\rm m} $
which appears after adimensionalization of the equation with the
maximal flow speed $U$ and a container typical scale
$L$. This necessary condition is useless to select flows which are
indeed efficient dynamos, i.e. flows with small critical magnetic Reynolds
numbers. Numerical kinematic dynamos are generally obtained from successive
attempts with flows given by a few velocity components selected for
analytical simplicity (see e.g. \cite{Dud89}). In this section, we present numerical 
results obtained with the mean experimental velocity fields presented above. 

As explained in paragraph 4.2.2, these velocity fields present a slight 
experimental disymmetry, which has a small influence on the obtained results.
We have thus prefered to first present the results with a symmetrized velocity 
field and then to study the influence of this additional parameter.
These results first concern threshold determination and the influence of different 
parameters on it. We then give a description of the spatio-temporal characteristics
of the self-excited magnetic field mode. In order to compare the numerical 
results to available experimental data, we finally study the response of the 
system to an external magnetic field for $ R_{\rm m} $ below the dynamo threshold.

\subsection{Threshold determination}

For a given experimental configuration, the mean velocity field shape is fixed,
and cannot be varied. As indicated in Section 2, varying the propellers 
rotation rate, we are only modifying $R_{\rm m}$ in the induction equation, 
but not the field shape. The effect of turbulence is not taken into account. 

For each experimental velocity field, we have performed a series of 
numerical runs with the kinematic dynamo code, at different $Rm$, 
and checked the energy evolution of each mode $m,n$, defined as follows:

$$E^{n,m}(t)= \int_0^1 | {\vec B}^{n,m}(r,t) | ^2 rdr \sim  e^{\sigma_{n,m} t}$$
Note that the different azimuthal $m$-modes are decoupled, since only axisymmetric
flows are considered.
Self excitation is achieved when the energy $E^{n,m}$ grows in at least a single
mode without external magnetic excitation that is if $\sigma_{n,m} > 0$ 
for at least a pair ${n,m}$. If $\sigma_{n,m} < 0\,\,\, \forall n,m$,  
ohmic diffusion dominates.

\begin{figure}[htbp]
\centering
\includegraphics[width=0.4 \textwidth,clip=true]{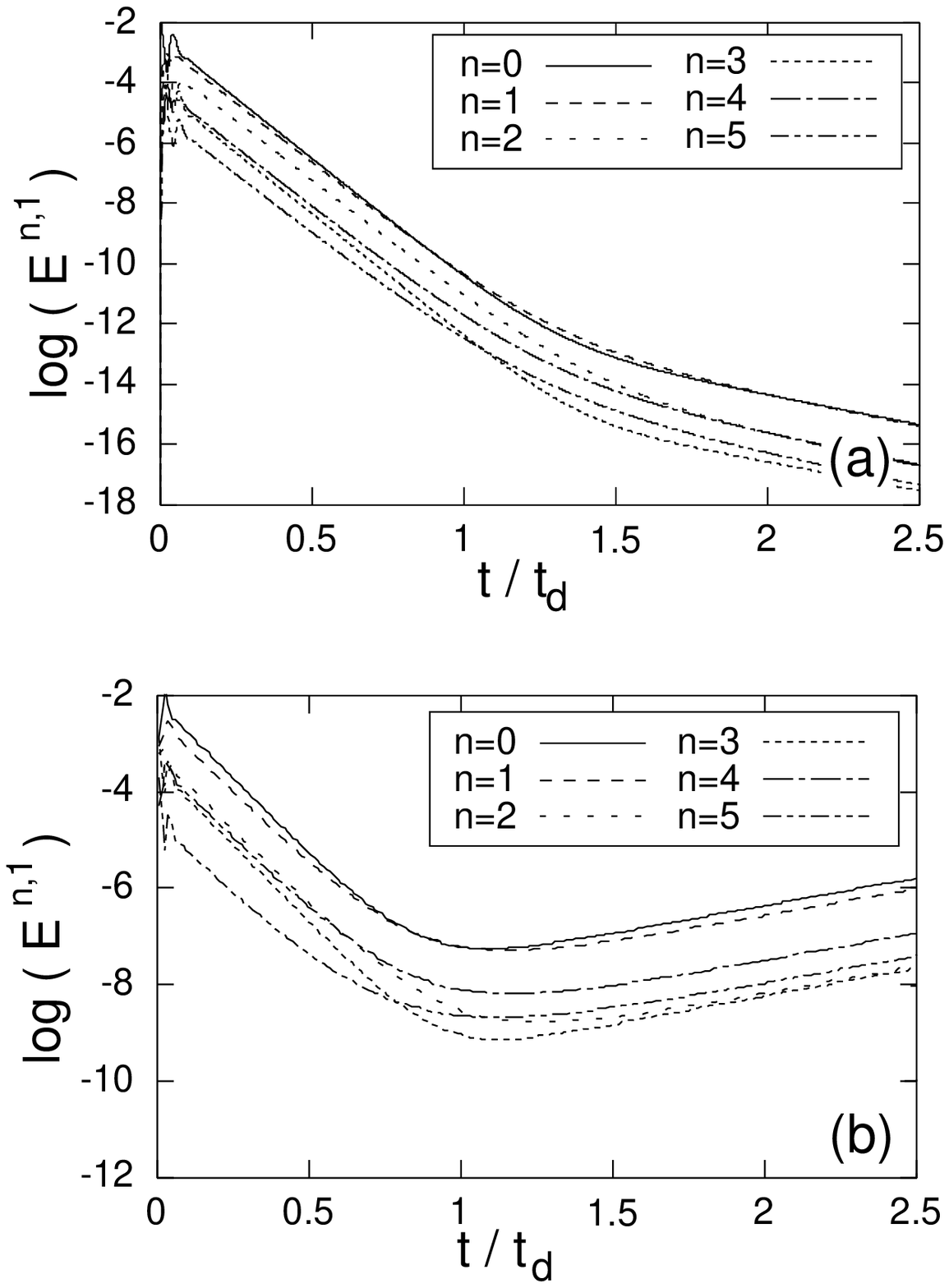}
\includegraphics[width=0.4 \textwidth,clip=true]{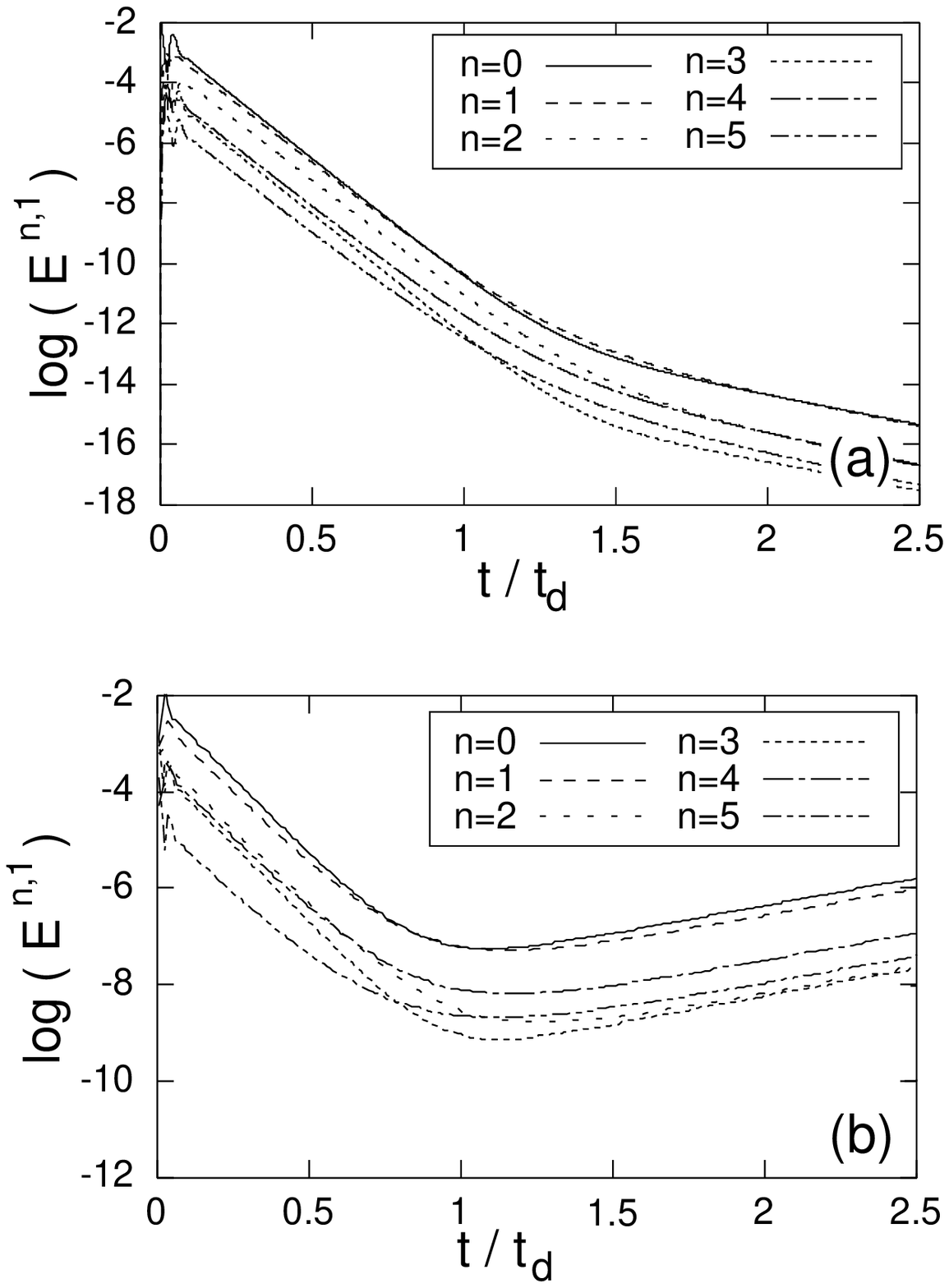}
\caption{Temporal evolution of the energy of modes $m=1; n=0,..,5$ for TM28 
propeller. (a) $R_{\rm m} = 100 < R_{\rm m}^c$ (b) $R_{\rm m} = 140 > R_{\rm m}^c$. 
The time unit corresponds to the ohmic diffusion time.
Two different regions can be distinguished: for $t/t_d \le 1$, relaxation of
the fast stable modes; for $t/t_d \ge 1$, relaxation of the slow stable modes (a), or
growth of the unstable modes (b). Note that in both cases the evolution is 
exponential with a growth rate $\sigma< 0$ (a) and $\sigma> 0$ (b).}
\label{Energy_t}
\end{figure}

Figure~\ref{Energy_t} displays the evolution in time of $E^{n,m}$ 
for $m=1$ and $n=0,...,5$ for  the velocity  field labelled TM28
under and above threshold.
The initial condition is:
$$E^{n,m}(t) = \left\{
\matrix{
1.0 &{\rm when\ } m = 0,1,2,3; \,\, n=0 & \cr
0.2 &{\rm when\ } m = 0,1,2,3; \,\, n=1,...,7 \cr
0.0 &{\rm in\ any\ other \ case} \cr
}\right.$$

In both cases, a transient regime corresponding to the relaxation of the 
initial field is observed until $t \simeq t_d$.
For $R_{\rm m} < R_{\rm m}^c$, the energy in all modes decreases in time, 
i.e. the growth rates are negative (figure~\ref{Energy_t}a).
For $R_{\rm m} \ge R_{\rm m}^c$, the energy of some modes begins 
to grow  (figure~\ref{Energy_t}b).
The critical magnetic Reynolds number $R_{\rm m}^c$ is defined 
as the value for which at least one growth rate is greater or equal to zero.
For the velocity fields tested, self-excitation always appears through  the
$m=1$ mode. Remember that, as the flow is axisymmetric,
an axisymmetric ($m=0$) self-excited magnetic field is forbidden by anti-dynamo theorems
\cite{Mof78}.

\begin{figure}[htbp]
\centering
\includegraphics[width=0.4 \textwidth,clip=true]{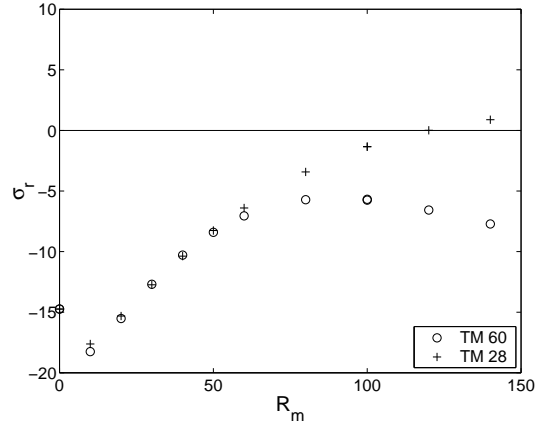}
\caption{Energy growth rates $\sigma_r$ for the most unstable mode ($n=1$ and $m=1$) 
as a function of $R_{\rm m}$. Crosses (resp. circles) correspond to 
the symmetrized velocity field of TM28 (resp. TM60) propeller. The TM28
propeller gives rise to dynamo action for $R_{\rm m}\simeq 120$ 
whereas TM60 does not.}
\label{Energy_sigma}
\end{figure}

In the following, for numerical efficiency, a different initial condition has 
been used, namely the (solenoidal) eigenmode of the vector laplacian which is closest
in shape to the most unstable mode of the induction equation (cf. section 4.2).
The energy is thus initially distributed on the $m=1$, $n$ odd Fourier modes. 
This leads to the near elimination of the transient regimes of figure~\ref{Energy_t}.
The variation of the maximal growth rate is presented in  figure~\ref{Energy_sigma}
as a fonction of  $R_{\rm m}$ for the two velocity fields of figure~\ref{MeanV}.
The velocity field corresponding to propeller TM28  exhibits a
growth rate that crosses the zero line for $R_{\rm m}^c \simeq 120$  
and gives rise to dynamo action. On the contrary, for the propeller TM60, 
the growth rate remains negative: it saturates for $R_{\rm m} \simeq 80$, and then 
decreases again. This latter result corresponds to a non-dynamo velocity field.
Performing simulations for $R_{\rm m}$ up to 300, we have observed that the growth rate
for TM28 (resp. TM60) keeps increasing (resp. decreasing).
At those large values of $R_{\rm m}$, simulation results are not accurate
enough to obtain a precise scaling of $\sigma(R_{\rm m})$.
A linear scaling of $\sigma$ with $R_{\rm m}$ would indicate
a saturation of $\sigma$, if expressed with the convective time unit $t_d/R_{\rm m}$.

The existence and the exact value of $R_{\rm m}^c$ strongly depends on the 
characteristics of the velocity field and on the boundary conditions. 
To investigate this dependence, we have modified different parameters.
First, the velocity fields have been modified in a controled manner in the 
numerical simulations, changing the ratio between the poloidal and toroidal 
velocity components. We have then investigated the effect of a conducting 
layer (with various thicknesses) surrounding the flow container.

\subsubsection{Ratio between the poloidal and toroidal velocity components}

\begin{figure}[htbp]
\centering
\includegraphics[width=0.4 \textwidth,clip=true]{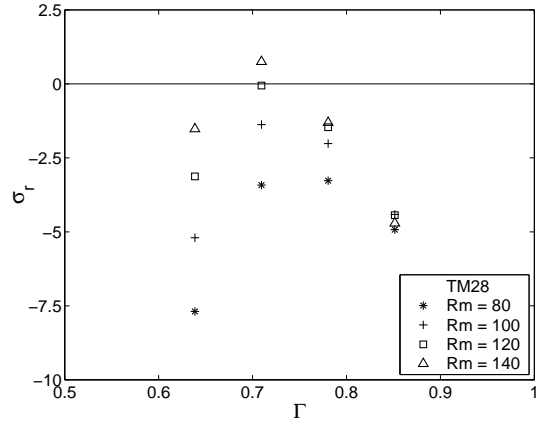}
\caption{Energy growth rates $\sigma_r$ for the mode $n=1$ and $m=1$ as a 
function of the mean poloidal-to-toroidal ratio 
$\Gamma= U_{pol}/ U_{tor}$ for various $R_{\rm m}$. 
The plotted data correspond to the symmetrized TM28 velocity field
which as a natural ratio $\Gamma = 0.71$.}
\label{Sigma_PolTor}
\end{figure}

Because of the flow axisymmetry (see section~\ref{velsection}), 
we can keep the solenoidal character of the flow, when changing the ratio 
$\Gamma = \overline { |{\vec U}_{pol}|}\,/\,\overline{ | {\vec U}_{tor} | }$
where
$$\overline {|{\vec U}_{pol,tor}|} = \int_V |{\vec U}_{pol,tor}| \,\,r\,dr\,dz $$ 
and look at the effect of $\Gamma$ on the critical magnetic Reynolds number. 
The $\Gamma$ parameter strongly affects the growth rates, as can be observed in
figure~\ref{Sigma_PolTor} for TM28 propeller: 
a negative growth rate can become positive and vice-versa.
Most of the studied velocity fields present a maximum growth rate for
$\Gamma \simeq 0.75$. This result is recovered for both dynamo and non-dynamo 
velocity fields such as the TM60. 
The optimal value $\Gamma_{\rm opt}$ corresponds nearly to the experimental value 
for the TM28 propeller ($\Gamma_{\rm exp}=0.71$), while it is different in the case of 
TM60 propeller ($\Gamma_{\rm exp}=0.82$). 

In a von K\'arm\'an flow, this parameter can be adjusted within certain limits
 in various ways e.g. by changing the diameter of the disc or the curvature of 
the blades (cf. Table~\ref{ComparativeVitesse}) or by fitting baffles located 
on the cylinder wall. Four baffles can be placed parallel to the cylinder axis 
at azimuthal intervals of $\pi/2$. These baffles break the axisymmetry of the problem, 
making it necessary to measure the full three-dimensional velocity field. 
This also imposes a higher resolution in the azimuthal direction in the 
code, making simulations more time-consuming. We have not performed systematic 
exploration of these effects, and will not present here the corresponding results.

Figure~\ref{Sigma_PolTor} also reveals a difference in the variation of the growth
rate with $R_{\rm m}$ depending on the value of $\Gamma$. For $\Gamma \simeq 0.5$,
$\sigma$ varies quasi-linearly with $R_{\rm m}$, while for  $\Gamma \simeq 1$, 
$\sigma$ seems independent of $R_{\rm m}$.

\subsubsection{Conducting layer}

It is empirically known that a stagnant layer of conducting material 
surrounding some fluid dynamos may reduce the critical magnetic Reynolds 
number. In the case of the Riga experiment~\cite{Gai00}, this has also been 
numerically verified by Stefani et al.~\cite{GaiXX}. The example of the 
Ponomarenko dynamo has been systematically examined in~\cite{Kai99}, varying the 
thickness of the static conducting layer, and the authors show that there 
is an optimal thickness leading to a lowest critical Rm.
A similar configuration is available in the VKS experiment~\cite{Bou02}
with a copper wall put inside the stainless steel cylindrical container. 

As  indicated in Section 3, the numerical simulations are performed
with a conductivity which is uniform inside a cylinder and insulating 
boundary conditions. A static conductive shell of  arbitrary thickness may thus be
easily introduced only if it has the same conductivity as the fluid. 
If the flow lies within a radius $R_c$, the stationary shell
goes from $r=R_c$, to $r=R_{ext}$, and we can use as control parameter
the relative width $W= (R_{ext}/ R_{c})- 1 $. Note that we have chosen 
to present the numerical results with the magnetic Reynolds defined with the 
radius $R_{c}$.

\begin{figure}[htbp]
\centering
\includegraphics[width=0.4 \textwidth,clip=true]{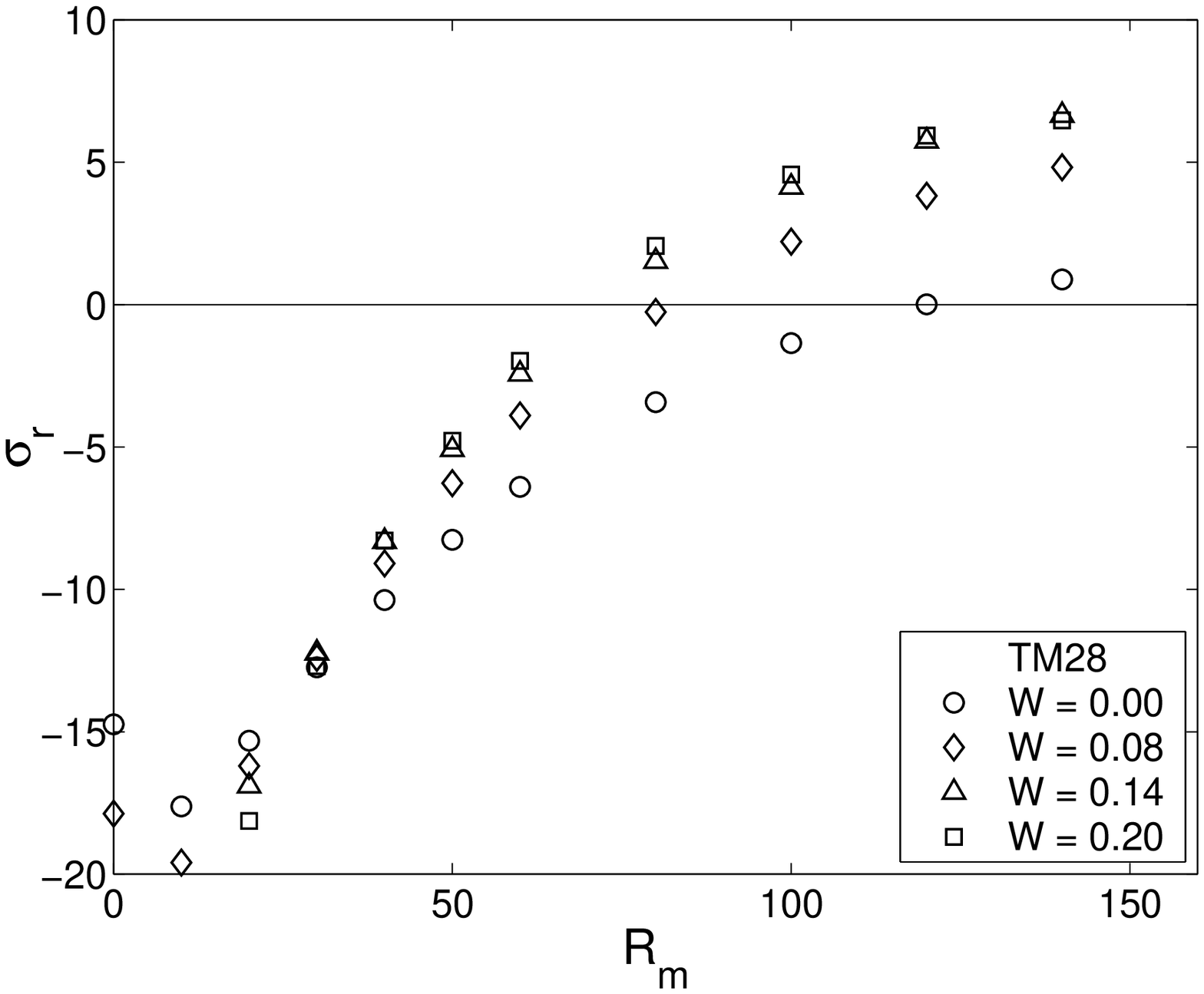}
\includegraphics[width=0.4 \textwidth,clip=true]{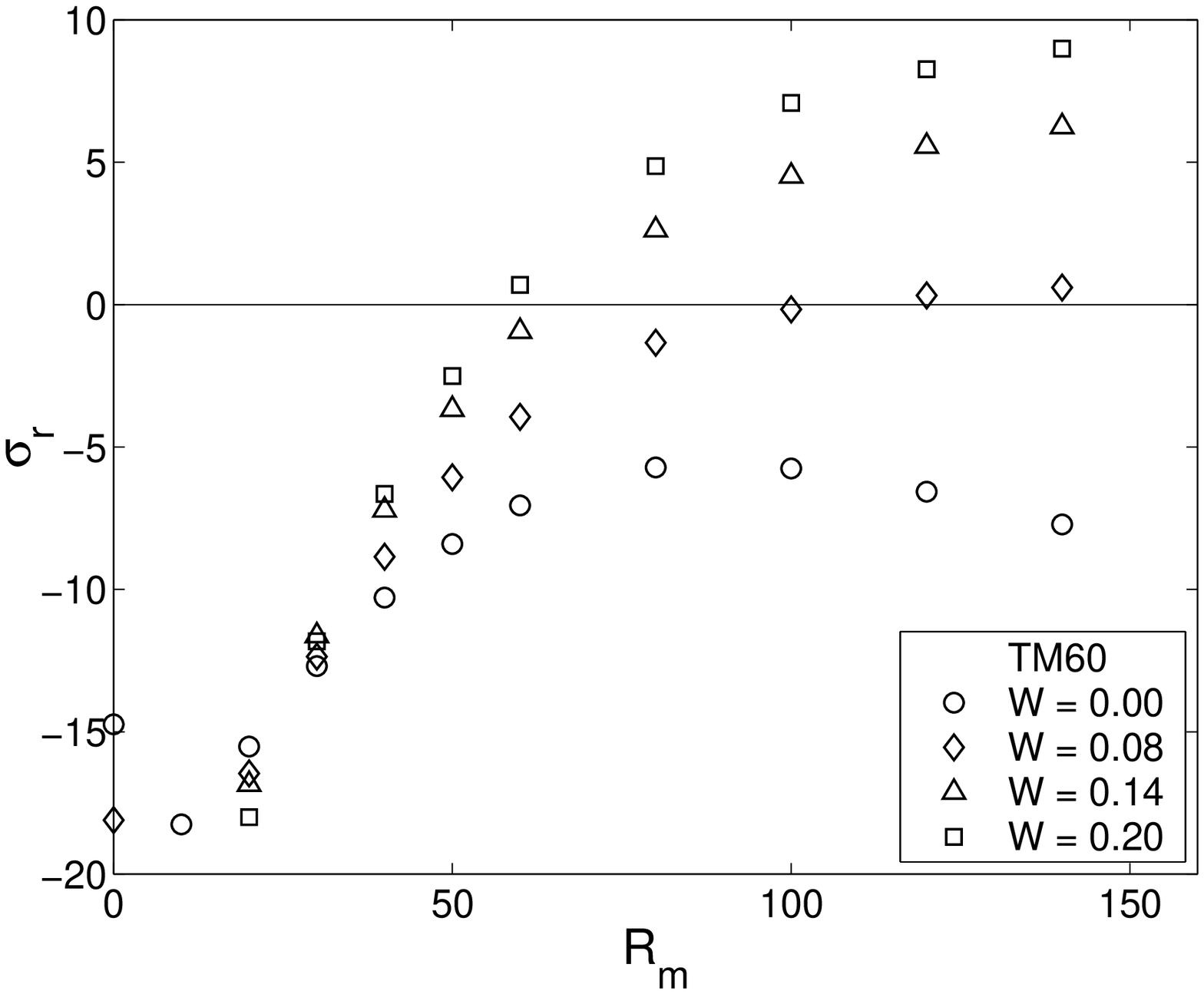}
\caption{Conducting layer effect: Maximal growth rate $\sigma_r$   
as a function of $R_{\rm m}$ for different layer thicknesses $W= (R_{ext}/R_c) - 1$;
symmetrized velocity field of propeller (a) TM28 and (b) TM60.}
\label{CondLayer}
\end{figure}

Figure~\ref{CondLayer} displays the effect of a conductive layer for propeller TM28
and for propeller TM60. 
The first effect concerns the growth rates that increase with $W$,
until they saturate for $W=W_s\simeq 0.2$. For propeller TM28, the threshold
$R_{\rm m}^c$ decreases from 120 for $W=0$ to 70 for $W=0.2$
(cf. Figure~\ref{CondLayer}a). 
The second important effect concerns the fact that the presence of a conducting
layer can actually change a non-dynamo velocity field into a dynamo velocity field: 
$R_{\rm m}^c =60$ for propeller TM60 and $W=0.2$, which is actually smaller
than the threshold for TM28 (cf. Figure~\ref{CondLayer}b).


\subsection{Description of the self-excited magnetic field}

In the following, we describe the self-excited magnetic field numerically observed with
propeller TM28. This field appears to have a very complicated spatial structure
and its temporal behaviour depends on the symmetry of the velocity field.

\subsubsection{Spatial characteristics}

It is well known that a smooth velocity field with a few spatial modes gives generally 
rise to magnetic eigenmodes with a broad band spectrum, and complex spatial configuration,
and we have verified that it is indeed the case for the flows we have used. 
Although small variations of the flow may induce large changes in the value of the 
critical magnetic Reynolds number, we have observed that the overall neutral mode 
topology is not much altered, so that we choose to present a single example.

As noted above, in an axisymmetric flow, the different azimuthal modes evolve 
independently and close to the critical $R_{\rm m}$,  it is the mode m=1 which has the largest
growth rate (recall that m=0 modes always decay). The magnetic axial modes (n) are
coupled by the kinetic ones, so that the axial spectrum is 
continuous from n=1 to the ohmic dissipation wavenumber.

\begin{figure}[htbp]
\centering
\includegraphics[width=0.4 \textwidth,clip=true]{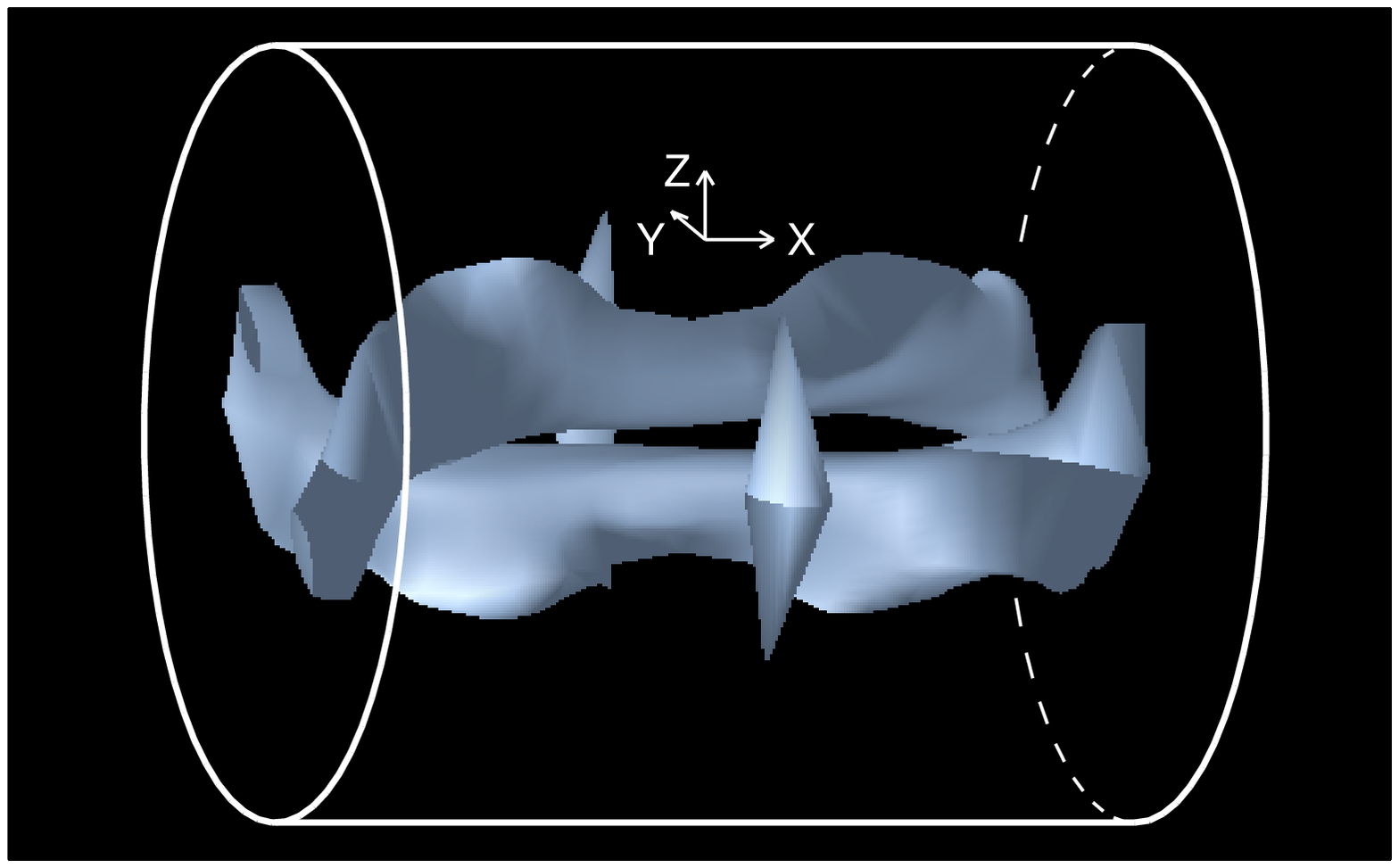}
\\
\hspace{-1.53 cm}
\includegraphics[width=0.315 \textwidth,clip=true]{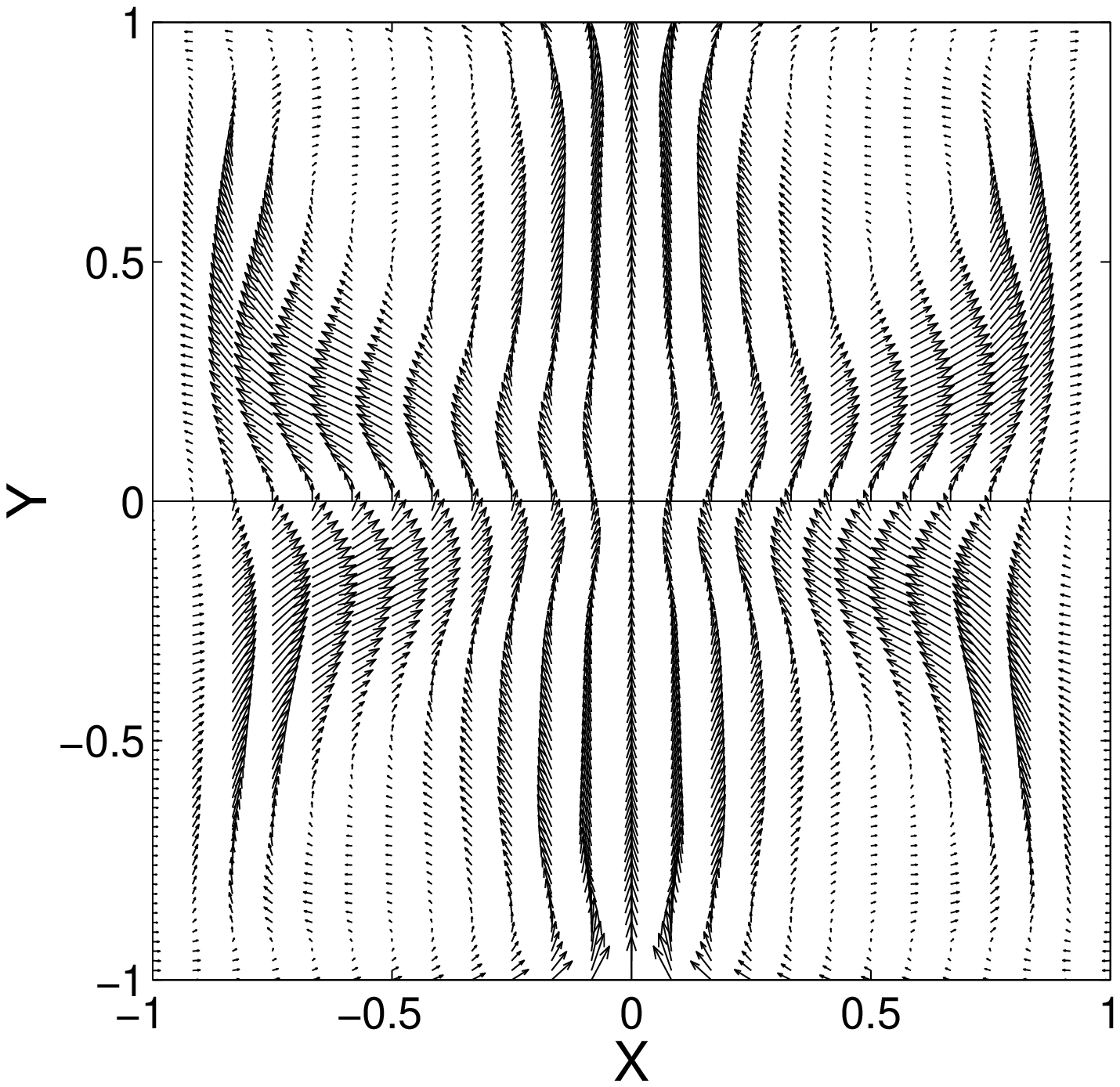}
\hspace{1.42 cm}
\includegraphics[width=0.315 \textwidth,clip=true]{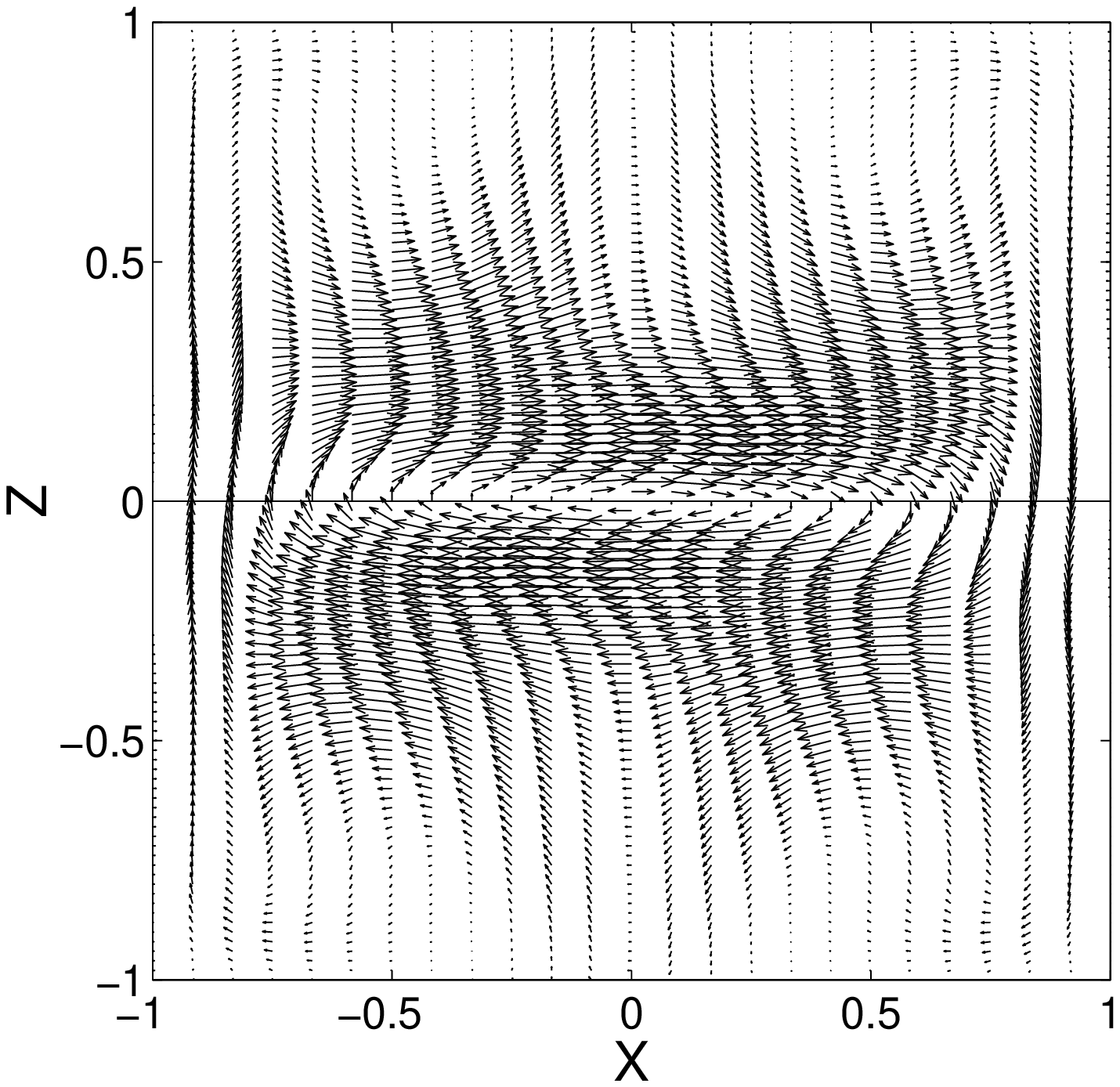}\\
\includegraphics[width=0.4 \textwidth,clip=true]{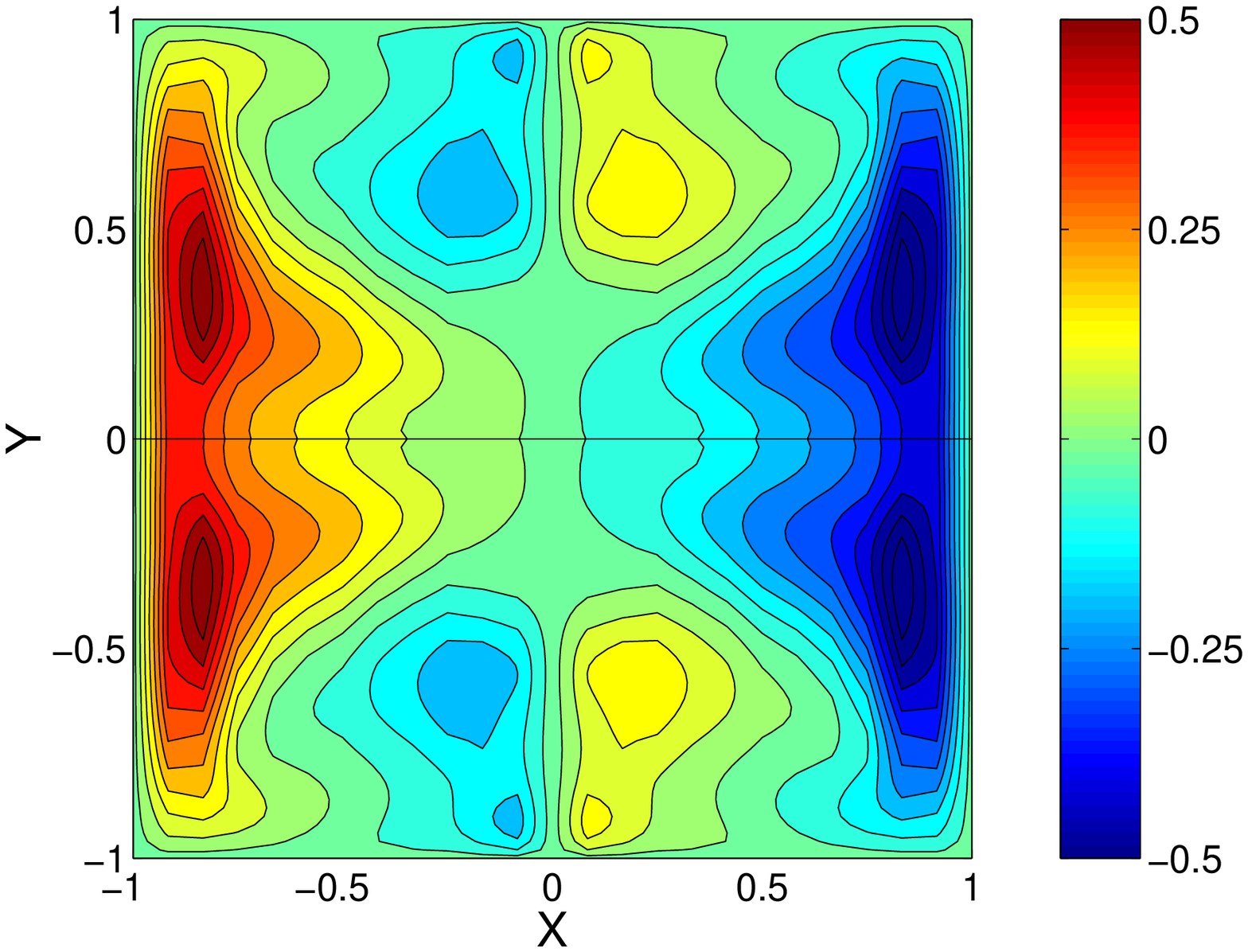}
\includegraphics[width=0.4 \textwidth,clip=true]{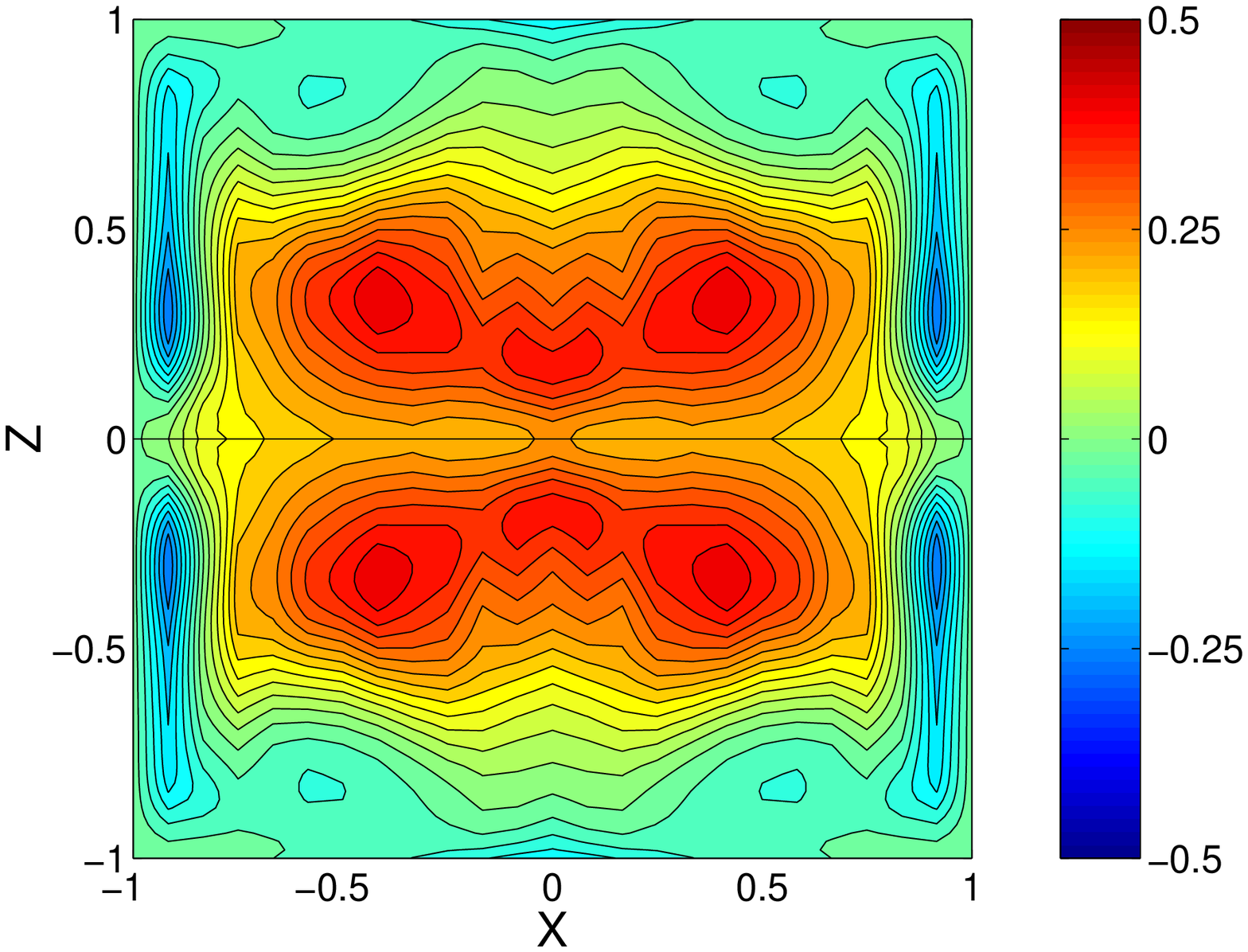}

\caption{Self-excited magnetic field of propeller TM28 for $R_{\rm m}=140$.
(a) Isosurface of the magnetic energy at 50 \% of the maximum value in the simulation
volume. Poloidal component of the magnetic field in (X,Y) plane (b) 
and (X,Z) plane (c).
(d) Z component of the magnetic field in (X,Y) plane; red (resp. blue) corresponds 
to vectors pointing out of (resp. into) the plane. 
(e) Y component in the (X,Z) plane; blue (resp. red) corresponds 
to vectors pointing out of (resp. into) the plane.}  
\label{B_Isosurface}
\end{figure}

Figure~\ref{B_Isosurface} represents a typical example of the growing mode
structure, obtained for $R_{\rm m} = 140$, with the TM28 velocity field,
in which symmetry has been artificially imposed. Figure~\ref{B_Isosurface}a 
shows the zones where the magnetic field intensity is higher than 50\% of 
its maximum value. The magnetic field appears to be strong principally in 
two banana-shaped regions, located on either sides of the cylinder axis.
Figure~\ref{B_Isosurface}b shows the poloidal component
of the magnetic field in the plane XOY of figure~\ref{Setup}.
In that plane, the magnetic field is roughly dipolar, oriented
perpendicularly to the cylinder axis. This corresponds to an m=1
angular dependency. On either sides of the dipole, we can see in 
figure~\ref{B_Isosurface}b the ends of the banana-shaped regions of 
figure~\ref{B_Isosurface}a. Figure~\ref{B_Isosurface}c shows the poloidal 
component of the magnetic field in the plane XOZ of figure~\ref{Setup}.
The magnetic field has a large amplitude in a significant portion of the regions
in this plane. The magnetic field there is axial, and has the expected m=1
angular dependency.
Figure~\ref{B_Isosurface}d and \ref{B_Isosurface}e show the components
of the magnetic field normal to the XOY and XOZ planes, respectively.

From all these figures, we can gather that the magnetic field of the most unstable
mode is roughly a dipole, oriented perpendicularly to the cylinder axis,
in the plane of figure~\ref{B_Isosurface}b, normally into the plane of
figure~\ref{B_Isosurface}e. Around this dipole, a group of
magnetic field lines comes upwards from the bottom left-hand of
figure~\ref{B_Isosurface}d, follows the arrows in the top half of 
figure~\ref{B_Isosurface}c,
and then goes downwards at the top right-hand of figure~\ref{B_Isosurface}d.
The m=1 angular dependency implies that another group of magnetic field lines goes
in the converse way, going into the bottom right-hand of figure~\ref{B_Isosurface}d,
to the left in figure~\ref{B_Isosurface}c, and then upwards in the top
left-hand corner of figure~\ref{B_Isosurface}d.

The electric current distribution associated to this magnetic field structure
is also quite similar in all situations. The electric currents concentrate in 
an elongated cylindrical region, located between the two regions of large magnetic 
field amplitude. Inside this region, the currents point perpendicularly to the axis, 
in the same direction as the magnetic field dipole.

\subsubsection{Temporal characteristics}

We have observed that the growth rate of the most unstable mode can have an 
imaginary part. The corresponding frequency is associated with a rotation of the 
magnetic field around the cylinder axis.
This frequency is very sensitive to the level of symmetry of
the velocity field.
In fact, for a very symmetric velocity field such as the TM28,
the neutral mode is nearly stationary, whereas for strongly dissymetric
ones the growth rate can have a much larger imaginary part.

In order to study this dependency, we have separated the TM28 velocity field
into two parts $\vec{U}_{even}$ and $\vec{U}_{odd}$ defined by their parities
with respect to the rotation around the OP axis of figure~\ref{Setup}.

In the case of TM28, we have checked that the velocity field
odd component $\vec{U}_{odd}$, being mainly due to experimental imperfections,
is small compared to the even component $\vec{U}_{even}$: 
$\vec{U}_{odd} \simeq 0.1 \vec{U}_{even}$.
We have then performed simulations of the composite flow
$\vec{U}=\vec{U}_{even}+\varepsilon \vec{U}_{odd}$ for various values of $\epsilon$,
while keeping the $R_{\rm m}$ based on $\vec{U}_{even}$ at a constant value of 160.
The value $\varepsilon =1$ corresponds to the experimental TM 28 flow.

\begin{figure}[htbp]
\centering
\includegraphics[width=0.4 \textwidth,clip=true]{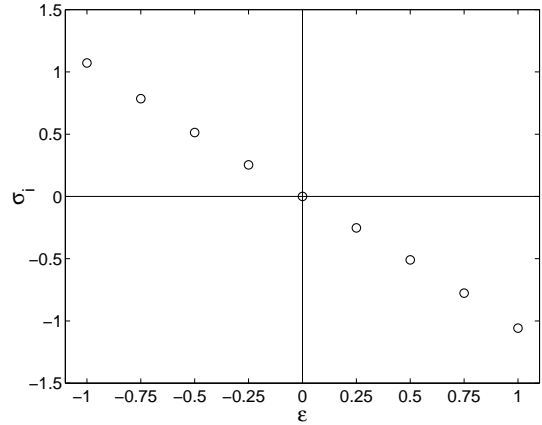}
\caption{Imaginary part of the neutral mode growth rate  $\sigma_i$ for $R_{\rm m}=160$
as a function of the symmetry parameter of the velocity field $\varepsilon$ 
as defined in the text. Symmetrized velocity field of TM 28 propeller which has a natural 
parameter $\varepsilon =1$, is defined on the even component of the velocity field. }
\label{Rotation}
\end{figure}

The dependency of the imaginary part of the growth rate on $\epsilon$
is shown in figure~\ref{Rotation}.
We can see that this dependency is very nearly linear, and
that $\epsilon=0$, which means a perfectly symmetric flow, is associated with a 
stationary magnetic field neutral mode.
This behaviour is observed with other velocity fields and seems to be robust in von
K\'arm\'an type flows. Note that the real part of the growth rate presents a
small variation with $\varepsilon$.

\subsection{Magnetic induction by an external field}
In the following, we want to address the following question: is it possible
to use the magnetic response of the experimental flow below $R_{\rm m}^c$
to an external magnetic field to predict the value of $R_{\rm m}^c$?

The kinematic code and the experimental
velocity fields can be used to obtain the response of the
system to an external magnetic field $\vec B_0$ and to perform 
a comparison with sodium experiments. We have checked the response to 
two different external magnetic fields: an axial field,
parallel to the axis of rotation (along $X$) or a transverse field, 
orthogonal to the rotation axis (along $Y$). 

In both cases, we have performed global and local measurements. 
We have first determined the decay times 
of the energy of each mode, and the saturation value of the
magnetic field induced inside the numerical box. In order to make a quantitative 
comparison with the results obtained in the VKS experiment, we have then studied
the local components of the induced magnetic field at a particular point in the
experiment (point P of figure~\ref{Setup}).

\subsubsection{Magnetic energy measurements}

Each numerical run is performed for a given velocity field as follows. The initial 
condition corresponds to no magnetic field inside the cylinder. At $t=0$, an 
external magnetic field is switched on and we let the system evolve in time.
This transverse field is directed along Y. It is sinusoidal with a $n=1$ axial dependency
and its maximal amplitude is 1.
We have chosen this external field structure because a uniform ($n=0$)
transverse field, which would have been easier to implement, as well as
closer to the experimental setup, is orthogonal to the growing magnetic field
eigenmodes we have observed. This feature is specific to the axially periodic
induction code.

\begin{figure}[htbp]
\centering
\includegraphics[width=0.4 \textwidth,clip=true]{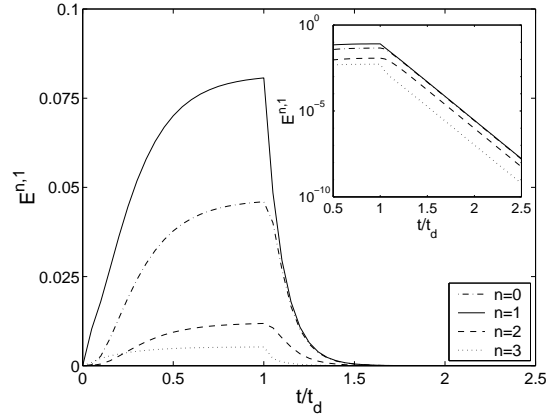}
\caption{Response to an external transverse magnetic field, varying sinusoidaly 
along the axis, of the symmetrized TM28 velocity field: 
temporal evolution of the energy for the  modes 
$m=1; n=0,...,3$ when $R_{\rm m} = 40 < R_{\rm m}^c$. The external 
field is applied for $ 0 < t < 1 $. It has an n=1 axial dependency, and has a 
maximal amplitude of 1.}
\label{ExternalB_t}
\end{figure}

\begin{figure}[htbp]
\centering
\includegraphics[width=0.4 \textwidth,clip=true]{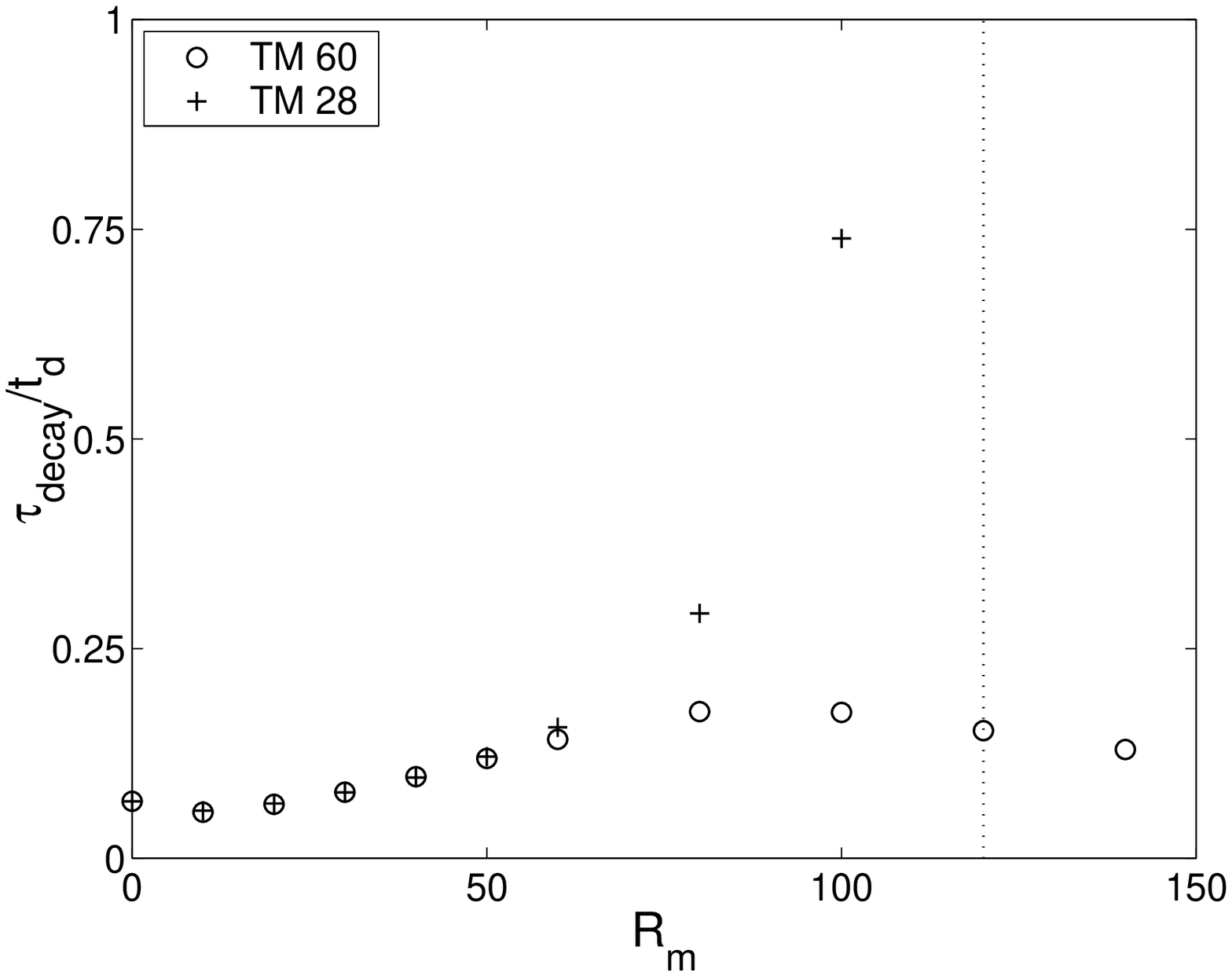}
\includegraphics[width=0.4 \textwidth,clip=true]{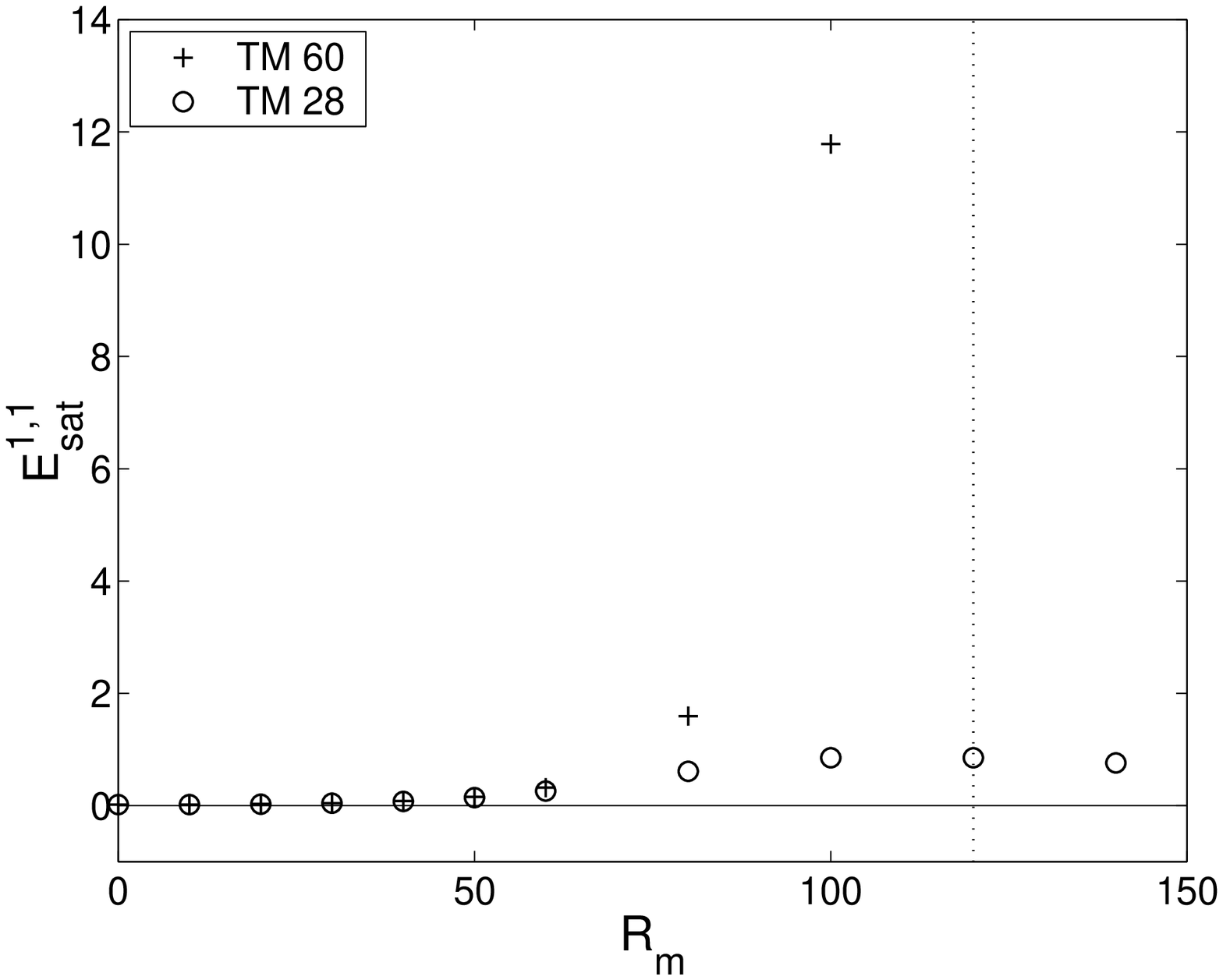}
\caption{variation of (a) the energy decay time $\tau_{decay}$ 
and (b) the saturation value of the magnetic energy 
$E^{n=1,m=1}_{\rm sat}$ as a function of $R_{\rm m}$ for an 
external transverse magnetic field  . 
Crosses: TM28 propeller; circles: TM60 propeller. Note that both
the decay time and the energy diverge when TM28 approaches threshold (dotted line).}
\label{ExternalB_Decaytime_Sat}
\end{figure}

For small $t$, the external field enters the cylinder,
until it approaches asymptotically a stationary saturated state 
of energy $E_{\rm sat}$ (see figure~\ref{ExternalB_t}, for $t<1$, $m=1, n=0-3$):
$$ E(t) = E_{\rm sat} \left( 1 - \exp ( - t / \tau_{\rm sat} ) \right) $$
Where $\tau_{\rm sat}$ is the saturation characteristic time. This time
diverges at the dynamo threshold of the TM28 system.
After the stationary state has been reached, 
the external field is disconnected, and we look at the magnetic energy  decay 
time (see figure~\ref{ExternalB_t}, for $t>1$):
$$E(t) = E_{\rm sat} \exp ( - t / \tau_{\rm decay} )$$ 
Figure~\ref{ExternalB_Decaytime_Sat} shows the variation of the decay characteristic
times $\tau_{\rm decay}$ and the saturation value of the magnetic energy
with $R_{\rm m}$. Both quantities  diverge for TM28 
propeller near $R_{\rm m}^c =120$ while they saturate near $R_{\rm m}=80$
and then decrease for TM60 propeller.   
We found that, below $R_{\rm m}=60$, theses features do not appear in practice to be 
good candidates to discriminate between a dynamo and a non dynamo velocity field.

\subsubsection{Local magnetic field measurements}

\begin{figure}[htbp]
\centering
\includegraphics[width=0.4 \textwidth,clip=true]{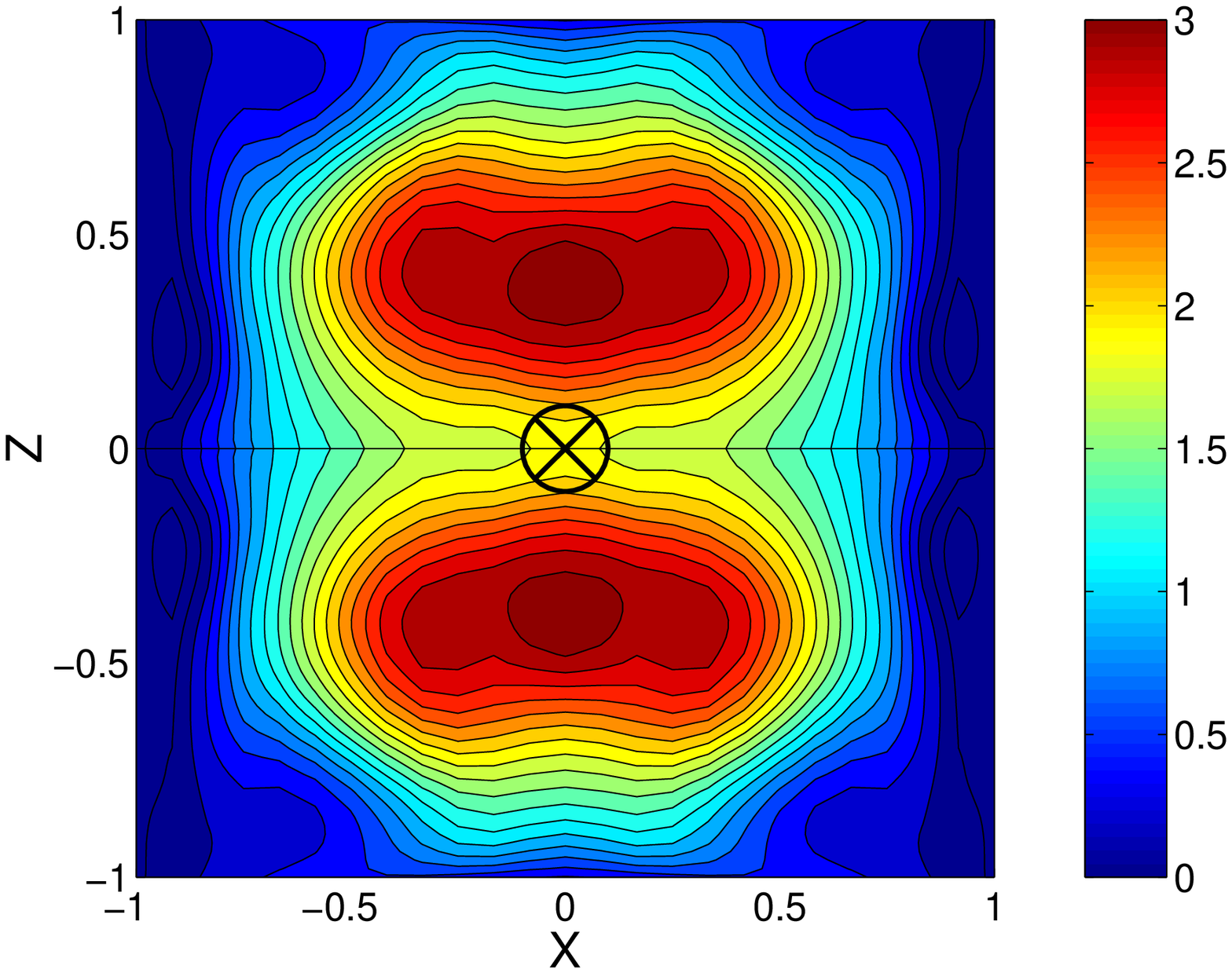}
\includegraphics[width=0.4 \textwidth,clip=true]{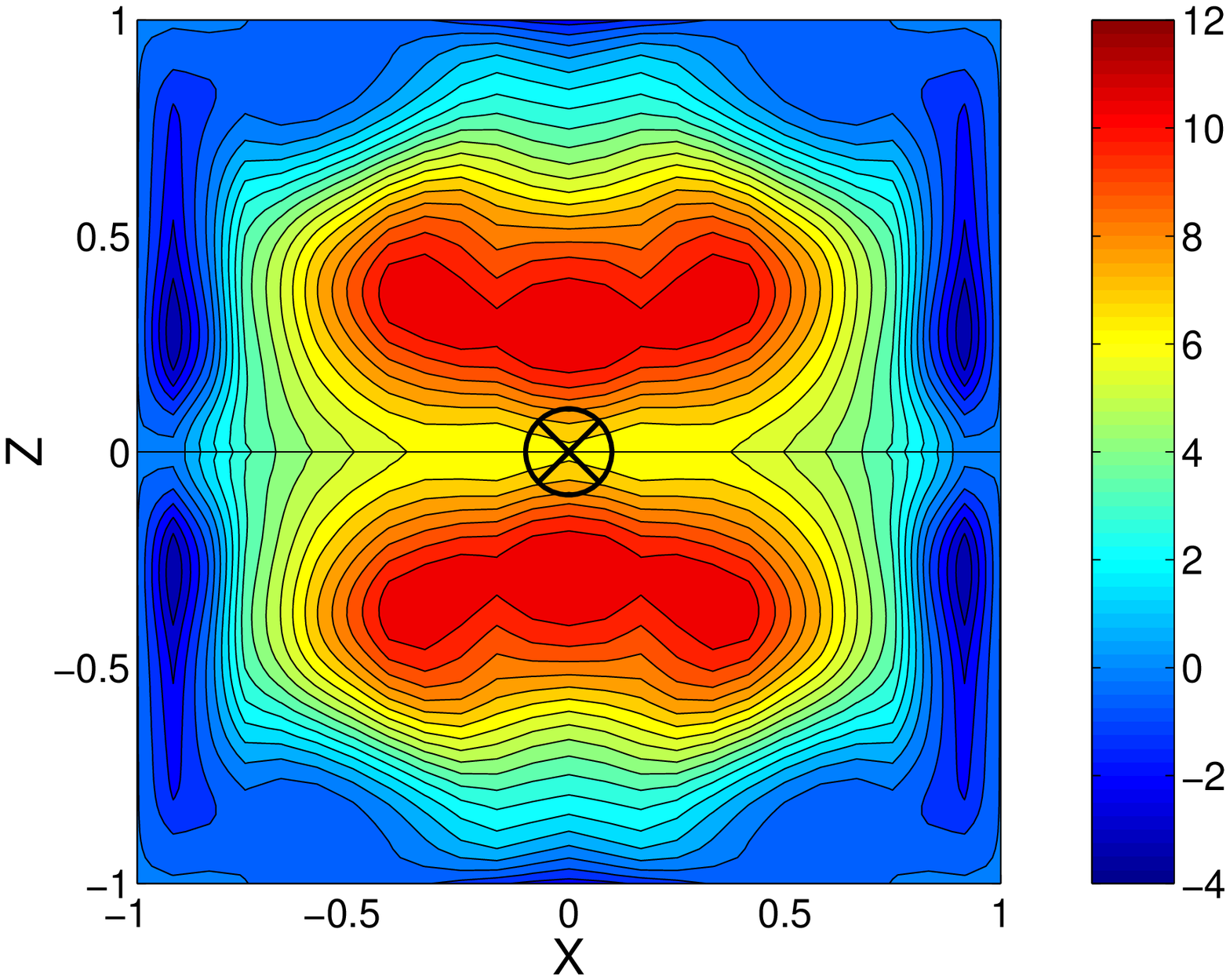}
\includegraphics[width=0.4 \textwidth,clip=true]{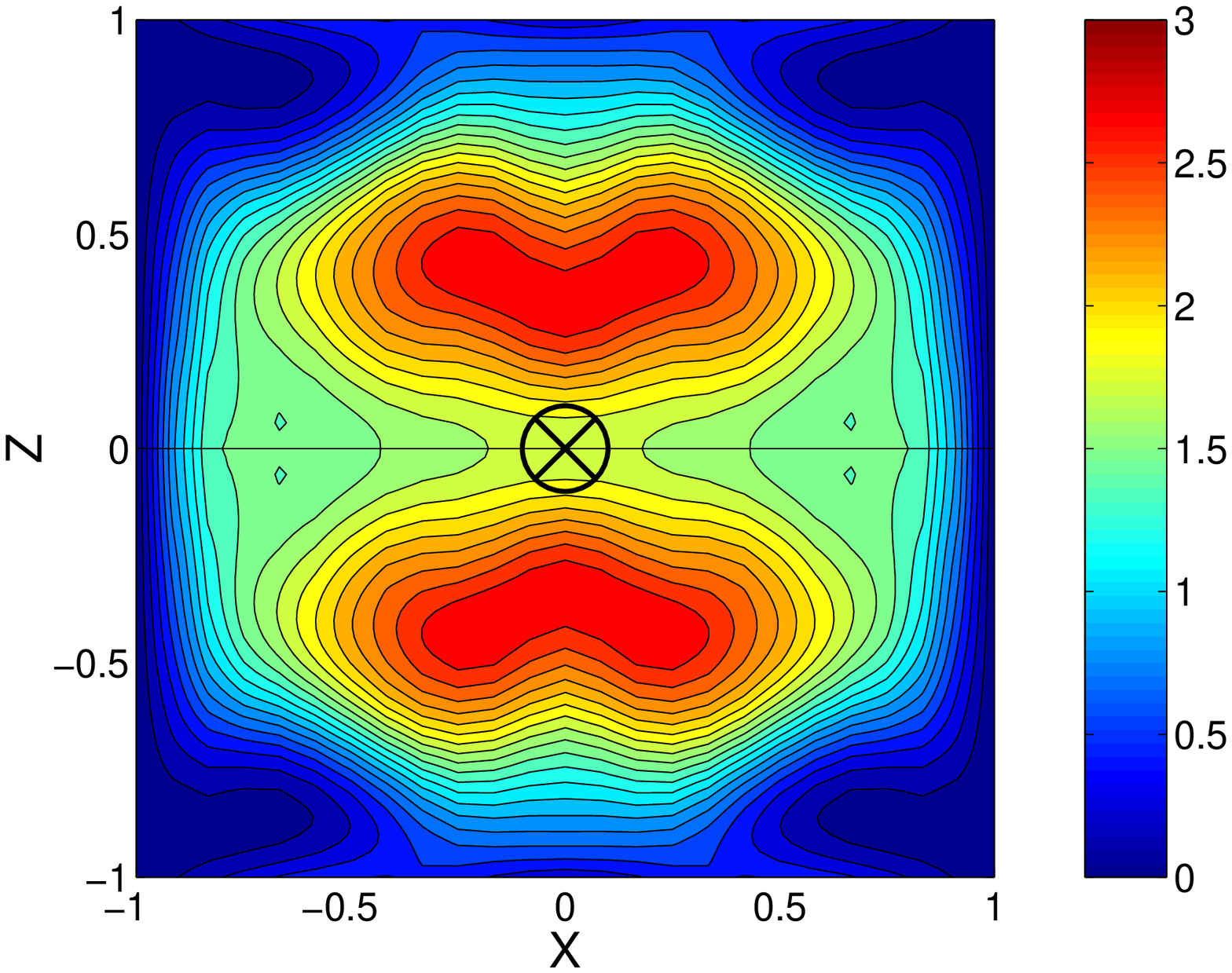}
\includegraphics[width=0.4 \textwidth,clip=true]{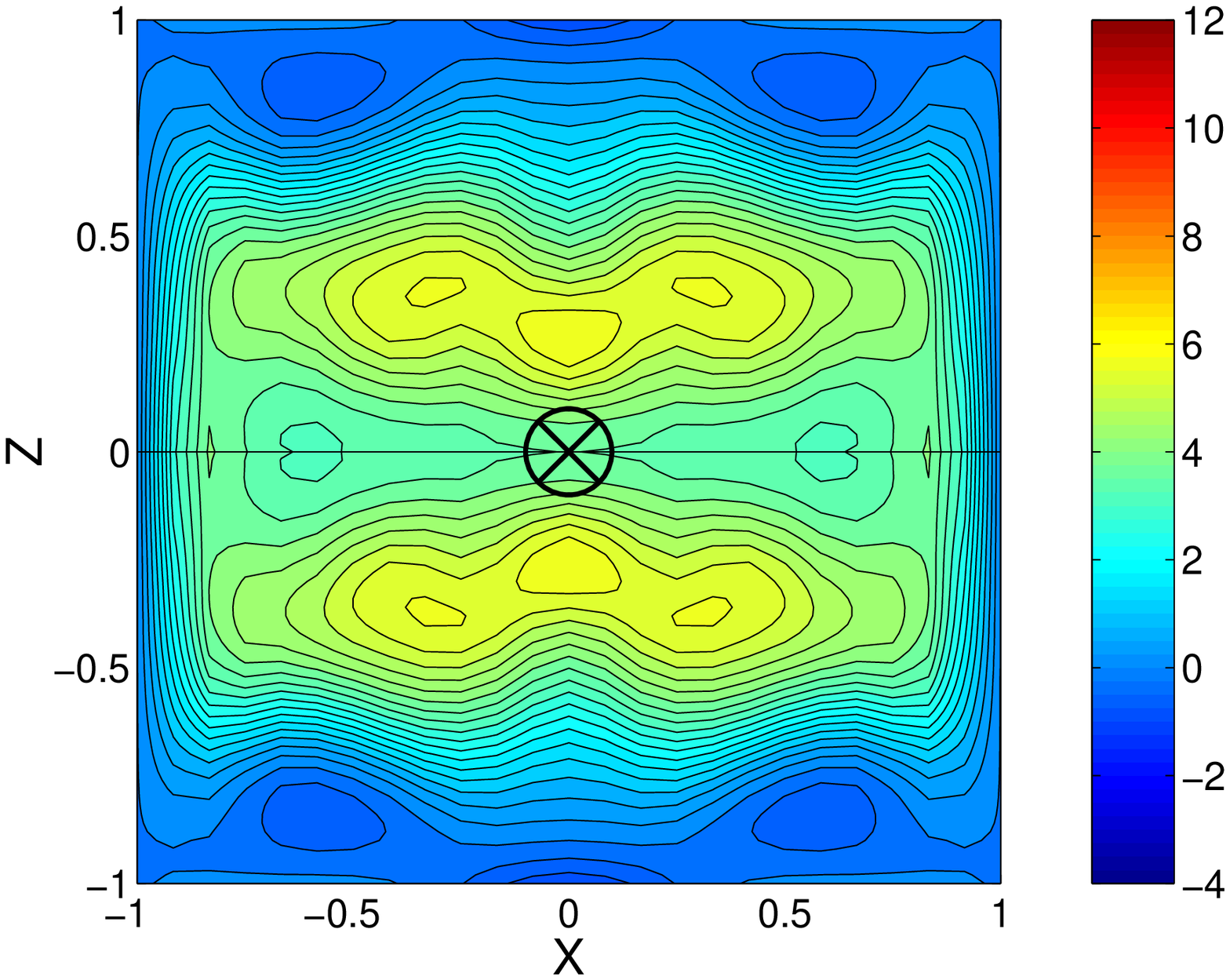}
\caption{Representation in the plane (X,Z) of the $B_Y$ component of the magnetic 
field in the presence of an external transverse magnetic field varying sinusoidally 
along the axis (reprented schematically by the central cross).
Propeller TM28: (a)$R_{\rm m}=50$ and (b) $R_{\rm m}=80$.
Propeller TM60: (c)$R_{\rm m}=50$ and (d) $R_{\rm m}=80$. 
blue (resp. red) corresponds to vectors pointing out of (resp. into) the plane }
\label{ExternalB_Coupes}
\end{figure}

The spatial structure of the magnetic field in the presence of the transverse magnetic field
is represented in the plane (XOZ) in figure~\ref{ExternalB_Coupes}.
For $R_{\rm m}=50$, the structures of the two magnetic fields look similar,
and the intensity of the induced field appears to be slightly larger for 
TM28 than for TM60 velocity field. The difference between the two fields  
increases with $R_{\rm m}$. For $R_{\rm m}=80$, the amplitude of the magnetic field corresponding to TM28
appears to be twice that corresponding to TM60. Spatially, a spreading of the magnetic
field is observed in the case of TM60, while strong gradients along $X$ can be 
evidenced for TM28, with an inversion of the magnetic field close to the disks. 

\begin{figure}[htbp]
\centering
\includegraphics[width=0.4 \textwidth,clip=true]{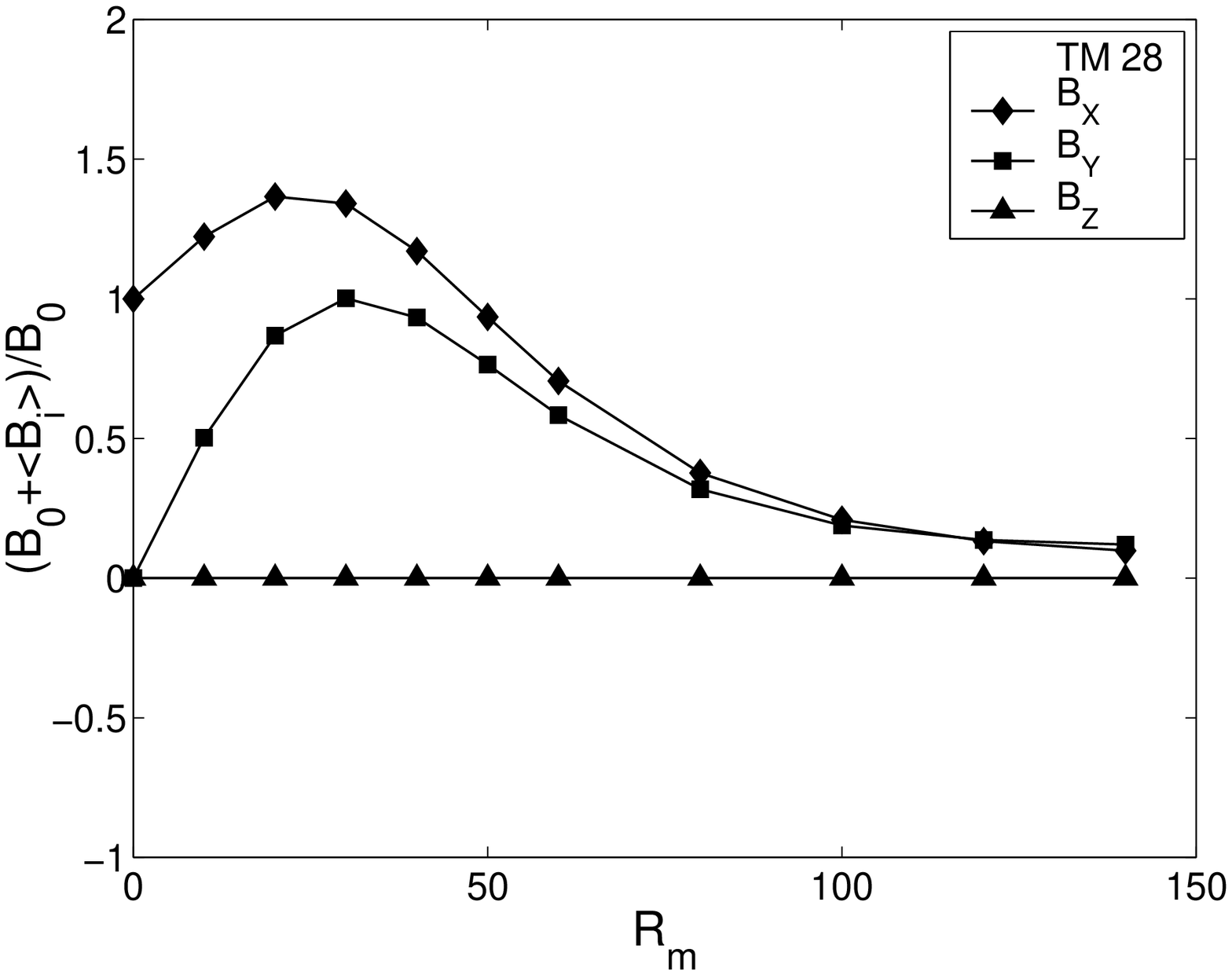}
\includegraphics[width=0.4 \textwidth,clip=true]{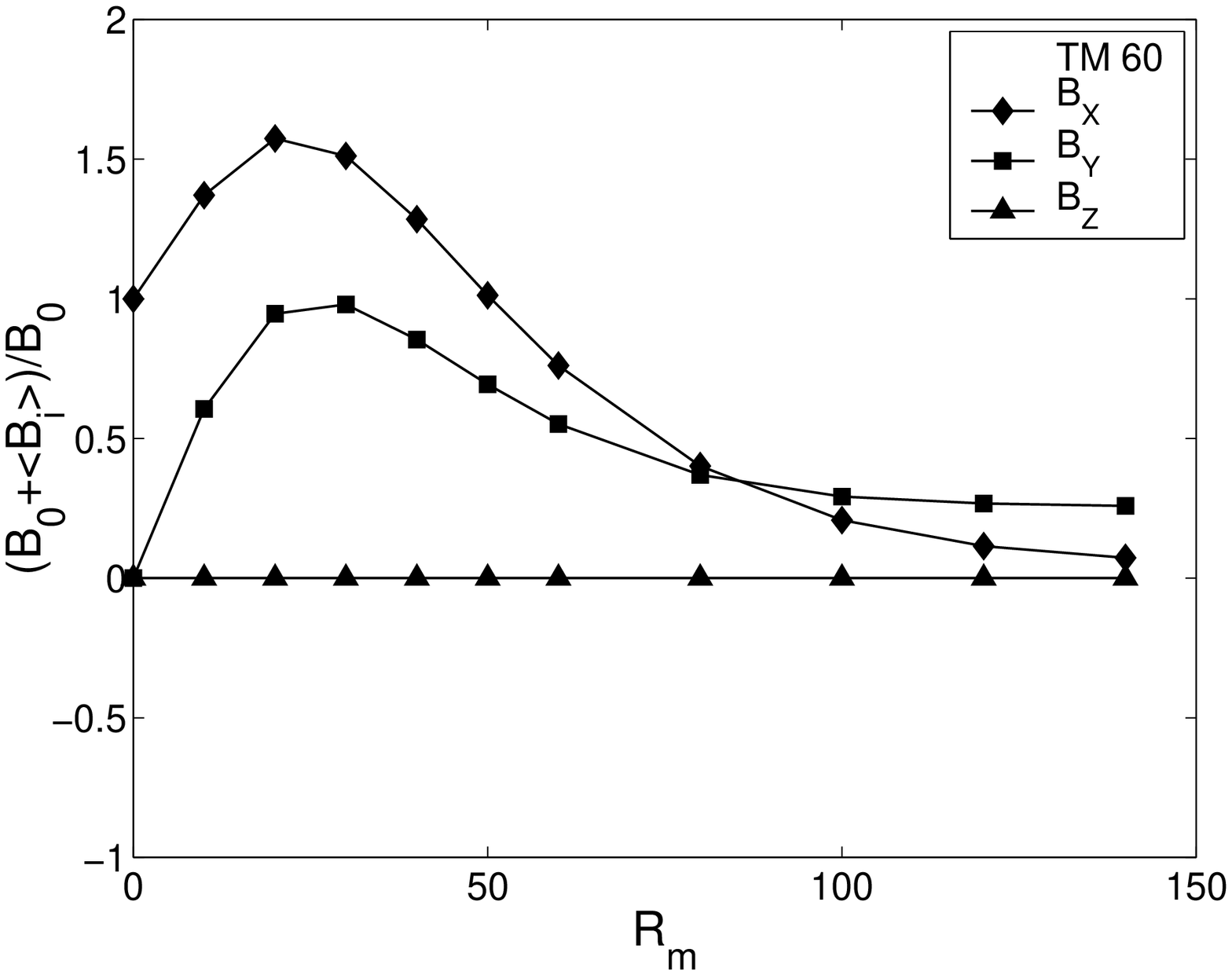}
\includegraphics[width=0.4 \textwidth,clip=true]{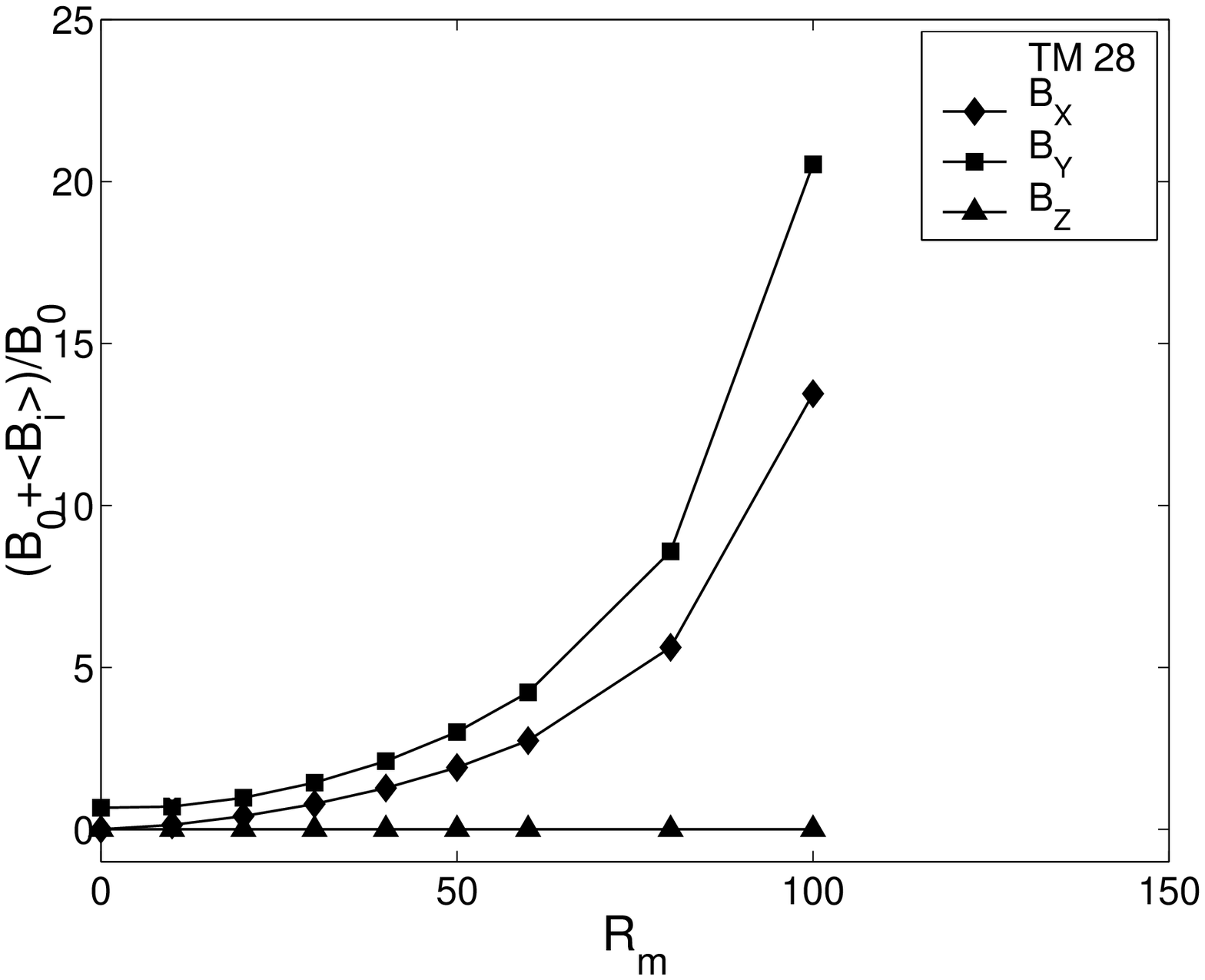}
\includegraphics[width=0.4 \textwidth,clip=true]{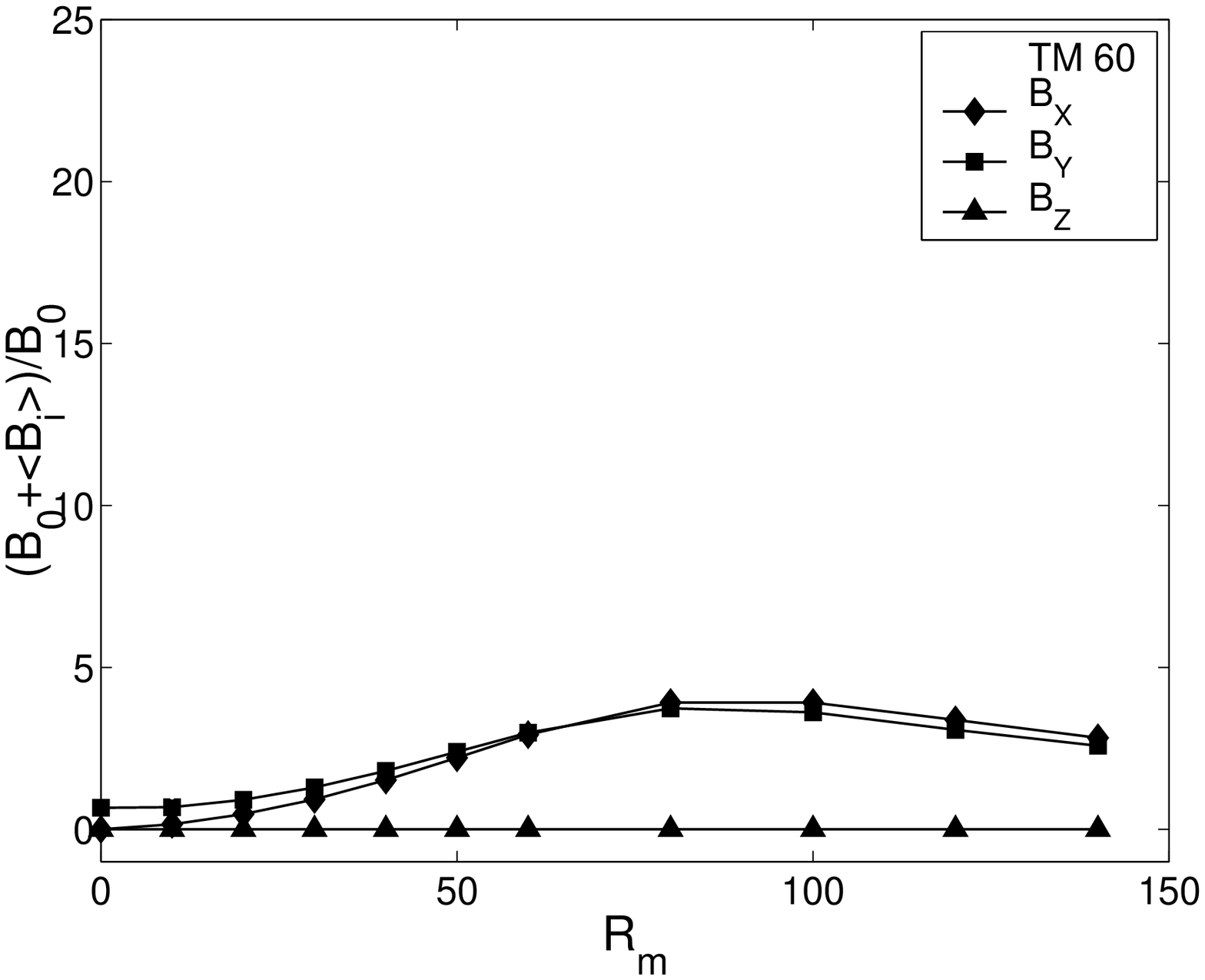}
\caption{Asymptotic values of three components of the magnetic field at position P as a 
function of $R_{\rm m}$. Diamonds, squares and disks respectively refer to $B_X$,
$B_Y$ and $B_Z$. Response of the TM28 (a) and the TM60 (b) symmetrized
velocity field, to an axial magnetic field.  
Response of the TM28 (c) and the TM60 (d) velocity field, to a transverse 
magnetic field varying sinusoidaly along the axis.}
\label{ExternalB_induction}
\end{figure}

We have studied the magnetic field induced at the point {\cal P} of
figure~\ref{Setup} by an external axial or transverse field in 
the symmetrized TM28 and TM60 flows. This point corresponds to the main
measurement location in the VKS experiment~\cite{Bou02}.
In figure~\ref{ExternalB_induction}, we have plotted the variation of the three 
components of the magnetic field as a function of $R_{\rm m}$.
We can see that in all cases the Z-component of the magnetic field is equal
to zero. We have checked that this is a consequence of the symmetry of the 
flow~\cite{Pet02}.

In the case of an axial external magnetic field 
(figure~\ref{ExternalB_induction} a and b), the
magnetic field components increase, then saturate for $R_{\rm m}\simeq 30$
and eventually decrease to zero for both propellers. 
This behaviour is akin to the expulsion of poloidal magnetic field expected in
poloidal circulation~\cite{Mof78} in spite of the existence of a toroidal flow. 

For a transverse field, figures~\ref{ExternalB_induction} c and d show that,
for the TM28 velocity field which gives rise to dynamo action, the
amplitude of the induced magnetic field components diverges as one approaches the
threshold. Conversely, in the case of the TM60 velocity field, which does not lead
to dynamo instability, the amplitudes remain bounded. Indeed we can see that the
magnetic field components increase for an interval in $R_{\rm m}$, then saturate
for $R_{\rm m}\simeq 80$, and eventually start to decrease. Inspection of the 
growth rates shows that the magnetic field growth rate, while remaining negative, 
has a maximum for the same value of $R_{\rm m}$ (cf. figure~\ref{Energy_sigma}).

\section{Discussion}

\subsection{Status of the von K\'arm\'an flow with respect to other dynamo flows}

Starting from an experimentally produced flow,  we have shown that it may 
produce kinematic dynamo action above some critical  $R_{\rm m}^c$, which could no be reached 
up to now in the VKS experiment due to insufficient driving power. 
It is instructive to examine how these 
results compare with  other known laminar axisymmetric dynamos, on the one hand, 
and available experimental dynamos, on the other hand.

Most numerical dynamo flows in a finite volume are produced in a spherical 
container, since the magnetic transmission conditions at the conducting-insulating 
interface are easier to implement.  For example,  the Von K\'arm\'an flow has the same 
symmetry properties as  the spherical configuration denoted $ s2t2 $ by  ~\cite{Dud89}, 
which has a $R_{\rm m}^c =54 $ with the particular radial function chosen by these authors. 
The same spherical $ s2t2 $ configuration has been chosen for the Maryland 
~\cite{Pef00,Pef01} and the Madison experiment  ~\cite{For01}, which use
two counter-rotating coaxial turbines set in motion by two motors to drive the flow.
The Maryland experiment uses a 30 cm diameter sphere and has not produced dynamo action.
The Madison installation is based on a one meter diameter sphere, where the velocity field 
could be measured using water as a fluid in a first step and the driving configuration 
has been optimized with the help of  a kinematic dynamo code to get $R_{\rm m}^c = 100 $. 
The liquid sodium installation is almost ready to be set in operation.

The two existing experimental dynamos in Riga and in Karlsruhe were based on 
flows of previous theoretical interest as occurrence of dynamo action in these flows may 
be proved analytically. They  are not axisymmetric but involve helicoidal streamlines 
winding around unbounded cylinders, the conducting fluid filling all space. In the 
experimental devices, the flow is constrained by cylindrical pipes of finite length, 
with guiding blades (Riga) or internal helicoidal walls (Karlsruhe) and occupies 
obviously a finite volume container.  In both cases, 
the design studies have concentrated on the search of a lower  $R_{\rm m}^c $. 
Although the two configurations are very different,  they lead to comparable critical 
magnetic Reynolds numbers based on a flow maximal speed V and a typical scale for 
the conducting fluid :
\begin{itemize}
\item Riga experiment : $R_{\rm m}^c \simeq 61 $, with $V= 15$ m/s,  $L= 0.4$ m ($L=$ 
external cylinder radius)
\item Karlsruhe experiment : $R_{\rm m}^c \simeq 36$, with $V= 4$ m/s,  $L = 0.9$ m ($L=$ 
container radius)
\end{itemize}

Recall from Section 4 that $R_{\rm m}^c $ around $ 60 $ (resp. $70$) has also been found 
for the TM60 (resp. TM28) flow with the adjunction of a conducting shell W= 0.2,
so that the lowest  $R_{\rm m}^c $ are indeed comparable for these cylindrical dynamos. 
There remains however a fundamental difference between these MHD flows at large 
$R_{\rm m} $: the turbulence level varies from a few percent for the flows with 
internal walls to 40\% for the VKS configuration, where the flow is not guided. 
In the latter case, it has been verified that the power scales as 
$R_{\rm m}^3 $~\cite{Bou02}, in agreement with dimensional analysis arguments. 
To reach the numerically predicted $R_{\rm m} ^c$, one has thus to overcome a power 
challenge: this is the price to pay to drive a flow without the constraint of internal walls,
such that its non linear saturation regime could have a link with those observed 
in natural dynamos.


\subsection{Threshold: comparison with VKS experiment}

The kinematic dynamo simulations based on experimental mean velocity fields of 
von K\'arm\'an flows exhibit the existence of self-excitation in some range of 
parameters that can be compared to experimental results of the VKS experiment.
The minimum threshold value $R_{\rm m}^c\simeq 60$ is found for the 
TM60 propeller, with a 20\% layer of liquid at rest. 
These numerical results correspond 
to a 4 cm sodium layer of conductivity $\sigma_{\rm Na}$ at rest, whereas 
the experiment is performed with a 1 cm thick copper boundary of conductivity
 $\sigma_{\rm Cu} \simeq 4 \sigma_{\rm Na}$.
Note that as far as the ohmic diffusion time is concerned, we expect 4 cm of
sodium to correspond to 2 cm of copper.

Using the following definition for the experimental magnetic Reynolds number:
\begin{equation}
R_{\rm m} = \mu_0 \sigma E_f R_c^2 2\pi f 
\end{equation}
$R_{\rm m}^c = 60$  would correspond to a critical rotation frequency $f^c=44$ Hz in the 
VKS experiment ($R_c=0.205$ m, TM60 propeller~\cite{Bou02}).
This last value can be  compared to the maximum frequency obtained in the sodium 
experiment $f=25$ Hz.  As the flow is highly turbulent, the power needed to maintain the flow 
scales as~\cite{Fri95}: 
\begin{equation}
P = K_P \rho L^2 U^3
\end{equation}
where $K_P$ is a dimensionless factor that depends on the geometry of the container
and of the shape of the propellers. We can write the magnetic Reynolds number as:
\begin{equation}
R_m =  \mu_0 \sigma E_f \left({PL \over K_P \rho}\right)^{1/3}.
\end{equation}
Going from 25 to 44 Hz thus implies to increase the power or the scale of the 
experiment by a factor 5, that is $P=750 $ kW or $R_c =1$ m.

In the case of the TM28 propeller, the minimum threshold value is
$R_{\rm m}^c\simeq 70$ which corresponds to a critical $f^c=41$ Hz. This lower value of the 
critical frequency of rotation can be related to the efficiency of  propeller TM28 
that is superior to that of TM60. In fact, the important parameter
of a flow configuration is given by the ratio $E_f / K_p^{1/3}$,
and a minimum $R_{\rm m}^c$ does not necessary corresponds to the most 
easily achievable critical frequency.


\subsection{Sensitivity to configuration parameters} 

\subsubsection{Propellers}

One aim of this work is to test the effect of various characteristics 
of the velocity fields on the existence and the value of
the critical magnetic Reynolds number $R_{\rm m}^c$.
One method would be to try different velocity fields that can be obtained from
analytical expressions, or to modify continuously an experimental one.
A continuous optimization could be performed in this way until
the velocity field with the minimum  $R_{\rm m}^c$ is obtained.
However, in a real system with no guiding walls, it is a technological
challenge to design a propeller able to reproduce the numerically optimized 
velocity field.

We have taken an experimental approach having in mind a possible sodium
experiment. Consequently, we have not varied a configuration in a continuous manner
but tried several driving configurations. Although the presented results concern only two
different propellers, we have tested approximatively 30 other configurations.
Various parameters have been varied: the height and curvature of the blades,
the propeller diameter, and other characteristics. Nevertheless, the presented 
cases correspond to representative ones, and enlighten the fact 
that two very close velocity fields can have very different dynamo properties.
The results do not reveal which characteristics a propeller must fullfill
to produce dynamo action, but nevertheless some conclusions can be drawn.

The poloidal to toroidal ratio appears to be a crucial parameter to obtain a 
dynamo velocity field, as already known from other numerical studies, in various
geometries.
The condition  $\Gamma \simeq 0.75$ (see figure~\ref{Sigma_PolTor}) corresponds
to a maximum growth rate for the 
explored velocity  fields, but it is not a sufficient condition.
No dynamo action has been observed with velocity fields with 
$\Gamma < 0.6$ or $\Gamma > 0.9$. The intuitive idea is then to obtain propellers 
with an appropriate $\Gamma$. This can be tried by changing the propeller geometry,
the curvature of blades
or by placing baffles in the inner part of the cylinder of the water experiment, i.e. parallel
to the generatrix. In this case, the toroidal component of the velocity field could 
be reduced, and, maybe, converted into poloidal velocity. The cost to be paid is 
that the velocity field is no more axisymmetric. In fact, the introduction
of baffles greatly changes the flow pattern, and the results are far more
complicated than expected: the poloidal component --- and consecutively $\Gamma$---
can even decrease with baffles. 

The relation of $\Gamma$ to the dynamo properties of the flow
may be appreciated with the help of the following toy model:
Assume that the toroidal velocity at scale $l$ is coupled to
the poloidal part of the magnetic field to increase the toroidal
field, as is the case in a pure differential rotation:
$$\frac{d\vec{B}_{tor}}{dt}=(\frac{u_{tor}}{l})\vec{B}_{pol},$$
and, conversely, for the poloidal component:
$$\frac{d\vec{B}_{pol}}{dt}=(\frac{u_{pol}}{l})\vec{B}_{tor}.$$
For a given total \lq \lq velocity\rq\rq  $u_{tor}+u_{pol}$,
the fastest growth rate (at scale $l$) is obtained when $u_{tor} = u_{pol}$,
i.e. $\Gamma = 1$.
This very crude argument based on two scalar variables may in principle
be examined further using numerical computations.

\subsubsection{Conducting layer}

In the section 4.1.2, we have observed that the magnetic energy growth rate
increases with a conductive layer. 
As the induction equation is a linear equation, the energy evolution
can be described by the equation $\partial_t E = \sigma_{n,m} E$ where
$\sigma_{n,m}$ is the growth rate of a given mode.
This growth rate comes from a competition between the magnetic field
generation, in some way proportional to $R_{\rm m}\, E$ and the ohmic diffusion.
This term takes into account the dissipation produced by the currents
$\vec j$ in the
conductive volume ($-\vec\nabla \times \vec j = \vec\nabla^2 \vec B$),
and then is proportional to $ E / L^2$, where $L$ is
the spatial scale of the diffusion. For a conducting shell
(with the same conductivity of the fluid) of size $L > R_c$, we can
write $L=R_c (1+W)$ and then the growth rate takes the form:
$\sigma \sim Rm - \frac 1 {(1+W)^ 2}$, as shown in figure~\ref{CondLayer}.
The effect of the conducting wall saturates, but the exact
values at which this effect is negligible depends on the velocity field,
and we are not able to obtain it.
This effect has been studied numerically in a recent work\cite{Kai99}.
This work shows that, for time dependent solutions, this effect can be perturbated by a skin
effect, that reduces the effective volume where the dissipation takes place.
In our case, as we are looking for a stationary  magnetic field (see Sec. 4.2.2), 
this skin effect does not appear.

\subsubsection{Symmetry}

In order to explain the temporal characteristics presented in  figure~\ref{Rotation},
we develop the following argments based on the symmetries of the flow.
When the flow is exactly even with respect to rotation of angle $\pi$ around the OP
axis of figure~\ref{Setup}, axial symmetry implies that it is even with respect to
rotation of angle $\pi$ around {\it any} axis going through O and perpendicular
to the cylinder axis.
If the magnetic field itself is even with respect to the rotation of angle $\pi$ around such
an axis, so will its time derivative $\partial_{t}\vec{B}$.
This means that if a magnetic field is even with respect to rotation around such an axis,
it will remain so. It is then straightforward to see that such a magnetic field must be stationary,
or consist of standing waves originating from the symmetry axis.

For a symmetrized velocity field and in the studied parameter range 
(i.e. $ \Gamma > 0.5$ and $R_m < 200$), the preferred 
structure for the magnetic field eigenmode is invariant with respect to rotation of
angle $\pi$ around the dipole axis. The magnetic field is then stationary at threshold.
If the flow does not have the required symmetry, the above argument does not apply.
The most general case is then that the magnetic fields will rotate around the cylinder
axis, hence be time-dependent.

We can check these arguments against the numerical results of Dudley and 
James\cite{Dud89}. In the case of their $s_2t_2$ flow, which has a structure 
close enough to that of a von K\'arm\'an flow, and possesses the required symmetry
properties, the magnetic field eigenmode is stationary at threshold.
Conversely, in the case of the $s_2t_1$ and $s_1t_1$ flows, in which no constraints
prevent the magnetic field from rotating around the axis of the flow,
the most unstable mode is always oscillating for $R_{\rm m} \neq 0$.

In the VKS experiment as in the present water experiment,
the symmetry can be broken very easily, and then 
generically we can expect a slightly rotating magnetic field, if the 
dynamo threshold is reached. Depending on the (slight) asymmetries, 
the frequency will be more or less important.


\subsection{Induction effects: comparison with the VKS experiment} 

The turbines that have been used in the first runs of the VKS experiment
correspond to the TM60 propellers used in this study. This should allow us to
compare the experimental data with the numerical results we have obtained,
and could provide us with a valuable check of the relevance of our
analysis process. Note however that this comparison can only concern
the mean values of the induction and that the very fast and large fluctuations 
observed in the VKS experiment~\cite{Bou02} cannot be taken into account.

\subsubsection{Induced magnetic field}

The accuracy of the threshold determination can not be checked,
since the threshold  expected with the TM60 turbines is beyond available power,
(see section 5.2).
At most can we say that the numerical results are not disproven.
Still, the response of the flow to an externally imposed
magnetic field has been measured in the VKS experiment,
and can be compared with the results of figure~\ref{ExternalB_induction}b
and \ref{ExternalB_induction}d.

\begin{figure}[htbp]
\centering
\includegraphics[width=0.4 \textwidth,clip=true]{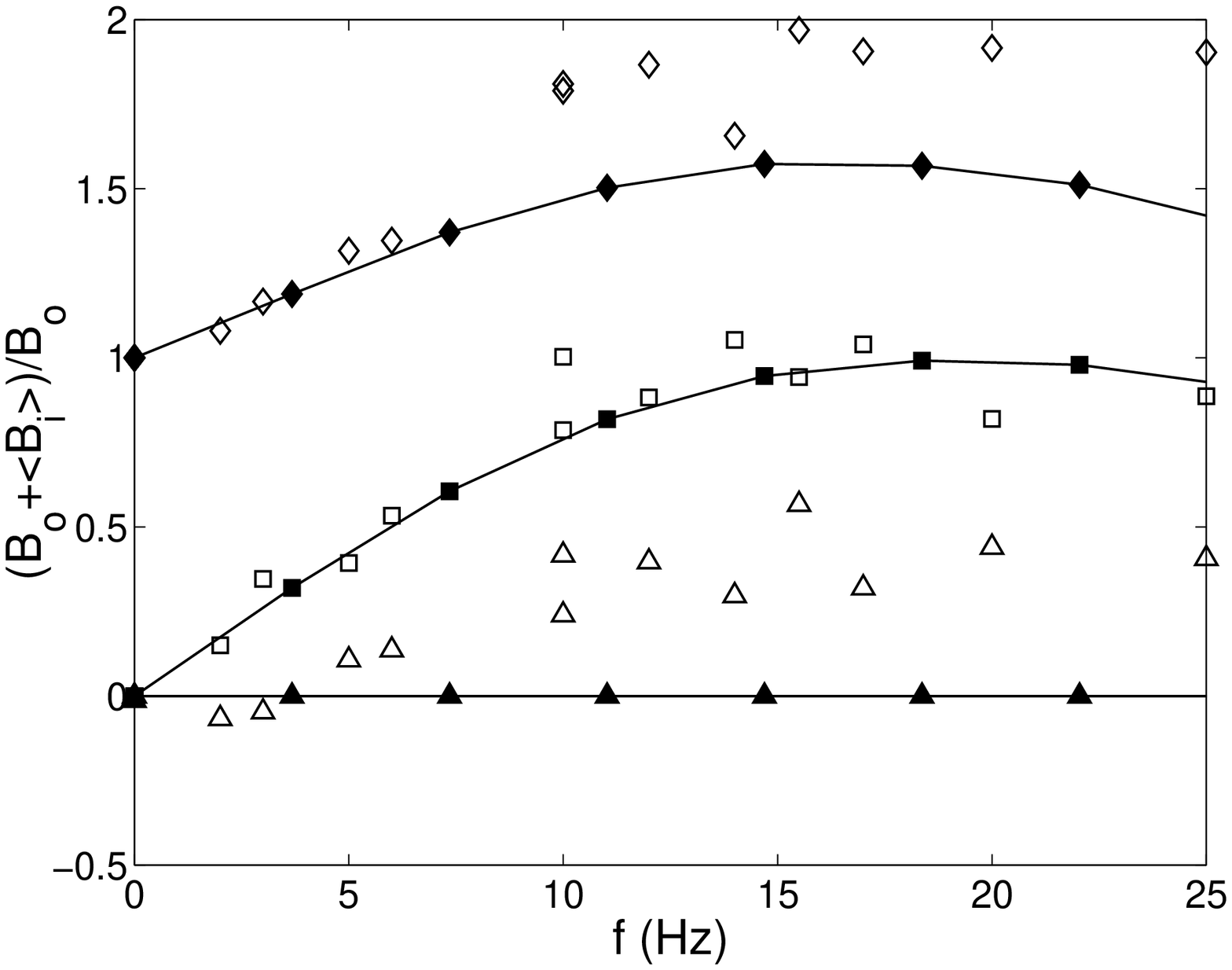}
\includegraphics[width=0.4 \textwidth,clip=true]{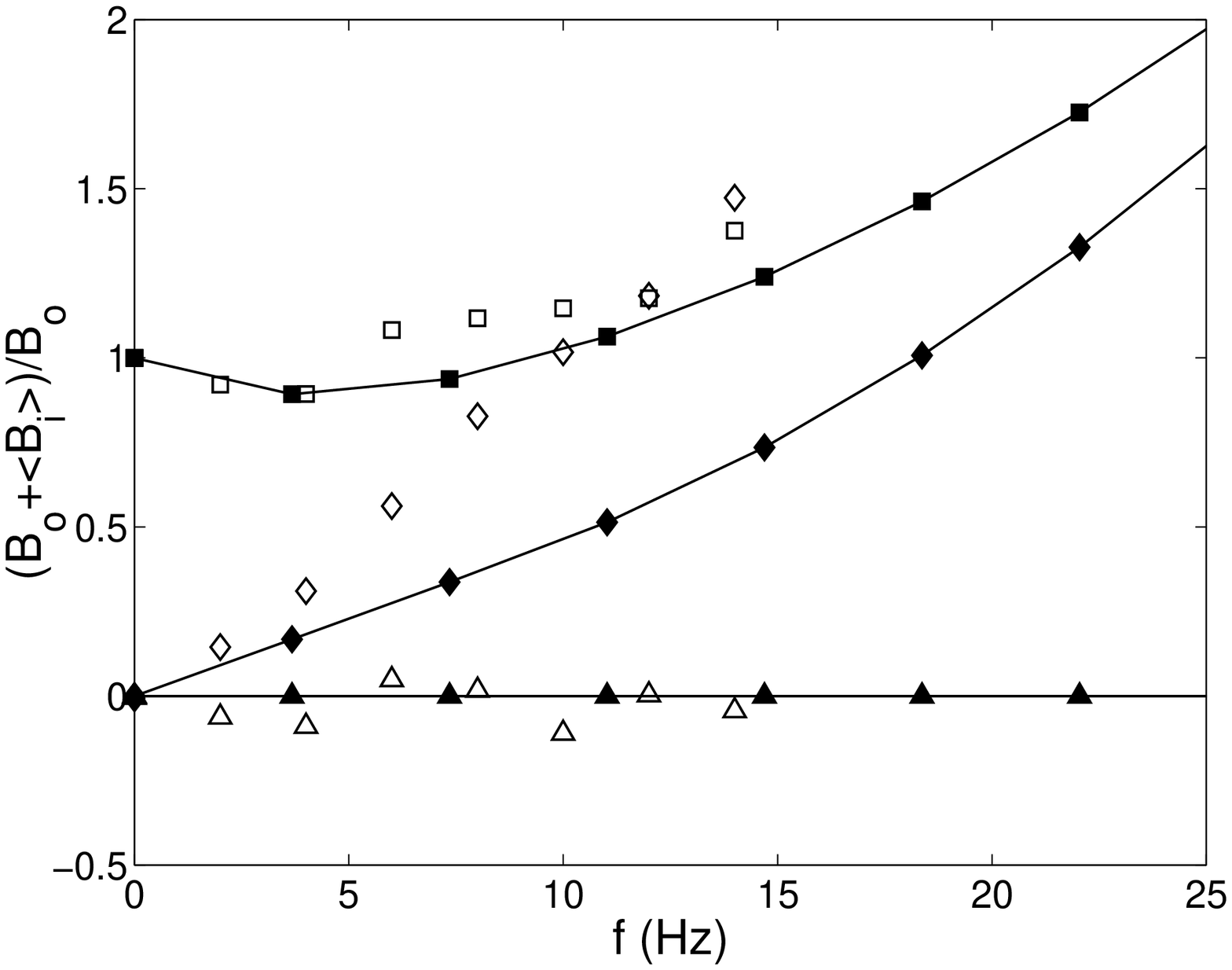}

\caption{Comparison with the VKS experiment in the case of 2 counter-rotating 
TM60 propellers. Response to an axial (a) and a uniform transverse (b) magnetic field.
Diamonds, squares and disks respectively refer to $B_X$,
$B_Y$ and $B_Z$. The dark symbols and solid lines correspond to the predicted 
behavior based on the velocity field used in the experiment. The white symbols
correspond to the measured data.}
\label{comparison_VKS_2D}
\end{figure}

Figure~\ref{comparison_VKS_2D}a shows the values of the magnetic field
components measured at the point P of figure~\ref{Setup}, with an
axial external field applied.
On the same plot are shown the numerical results obtained for an applied field
that is axial and uniform outside the cylinder and with no conducting layer.
We can see that the measured values of the Y-component
agree fairly well with the simulation results. The agreement is correct but 
not as good for the X-component of the magnetic field.
Lastly, we can see that the Z-component of the measured
magnetic field is significantly different from zero,
its expected value. This seems to imply that the symmetry considered in
paragraph 4.3.2, which forces the Z-component to zero, is somehow broken in
the experiment.

The results obtained for a transverse external field are presented in
Figure~\ref{comparison_VKS_2D}b. The transverse external field is this time
uniform outside the cylinder, with no conducting layer. We can see a strong 
discrepancy between the measured and the simulated values of the X-component.
Indeed, the measured values seem to be a factor of two larger than the computed
results. On the contrary, the agreement is good for the Y-component.
Finally, we can see that the measured Z-component remains in this case
close to zero: either the symmetry is no longer broken
in these experimental runs, or the configuration is less sensitive to dissymetry.
Numerical runs have also been performed with a $W = 0.2$ (see paragraph 4.1.2) conducting layer,
showing only slight differences.

\begin{figure}[htbp]
\centering
\includegraphics[width=0.4 \textwidth,clip=true]{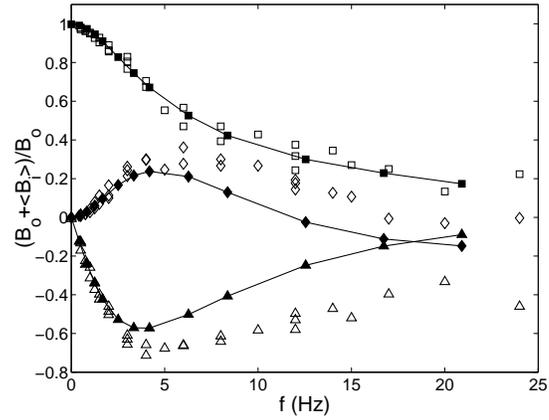}
\caption{Comparison with the VKS experiment in the case of one rotating TM60 disk.
Diamonds, squares and disks respectively refer to $B_X$, $B_Y$ and $B_Z$. 
The dark symbols and solid lines correspond to the predicted
response to a uniform transverse magnetic field based 
on the velocity field used in the experiment. The white symbols correspond
to the measured data.}
\label{comparison_VKS_1D}
\end{figure}

Finally, figure~\ref{comparison_VKS_1D} shows the results obtained in the case
where only one TM60 turbine is rotating inside the vessel, the second being at rest,
with a transverse field applied.
The agreement is this time far better than in both the above
configurations, at least for small values of the rotation rate.
The Y-component of the magnetic field decreases
to zero. This corresponds to an expulsion of the applied transverse magnetic
field~\cite{Wei66,Mof78,Odi00}. The X- and Y-components of the induced field 
also exhibit  a quadratic dependency on the rotation frequency. 
Finally, we have checked numerically that 
the velocity field obtained for one
rotating disk does not give rise to dynamo action for $R_{\rm m} <200$.

To sum up, we can see that the agreement between simulation and measurements
can be satisfactory in the case where only one disk rotates, or
less satisfactory in the case where both disks rotate.
The good agreement in the case of one disk 
seems to imply that the numerical code and the measurement process are
relevant. In the case where two disks rotate, however,
we can see that the measurement point lies precisely at the position where the
turbulence intensity is highest, and the average flow magnitude is weakest.
It may be possible that the time-averaged velocity field is
not sufficient to properly reproduce the experimental induction effects at this position.

\subsubsection{Decay rates}

It would be interesting to check the decay times presented in 
figure~\ref{ExternalB_Decaytime_Sat} against the experimental value.
For the maximum value of the magnetic Reynolds number achievable 
in the VKS apparatus, the magnetic field 
energy decays in $\simeq 0.12 \mu_0 \sigma R_c^2 \simeq 0.05$ s for TM60 
propeller with no conducting layer. Propeller TM28 exhibits the same decay time 
(within 2\%) at $R_{\rm m} = 50$. This means that, in an experiment without 
conducting layer, it is very difficult to determine by pulse decay measurements
alone if the turbines are of the TM28 ({\sl i.e.} capable of dynamo action)
or of the TM60 ({\sl i.e.} incapable of dynamo action) type.

If a very large conducting layer ($W = 0.2$) surrounds the experiment, however,
the magnetic field energy decays in $\simeq 0.12$ s for TM28,
and $\simeq 0.23$ s for TM60.
In this case, it would be possible to determine which velocity field has 
the smallest threshold by pulse decay measurements.

\subsubsection{Turbulent effects?}

In the VKS apparatus, the flow shows very large turbulent fluctuations,
which in turn induce very large magnetic field fluctuations \cite{Bou02}.
It is expected that under certain circumstances, such fluctuations
could induce a large scale component of the magnetic field through what is
termed an $\alpha$-effect~\cite{Mof78}.
We have shown (figure \ref{comparison_VKS_1D}) that, if only one
propeller rotates, our numerical results are quite close to the experimental data,
though the numerical simulations have been performed with the time-averaged component
of the flow only. This good agreement could imply that, 
in that case, the leading contribution to the
induction effects comes from the time-averaged component of the flow.
Conversely, a discrepancy is observed in the case of two counter-rotating disks,
where the time-averaged component of the flow is weaker, and where large vortices 
sweep the measurement location. This could possibly be ascribed to turbulence effects, 
either $\alpha$-effect or reduction of the electric conductivity~\cite{Mof78,Rei01}.

\section{Conclusion}

The existence of dynamo effect has been recently confirmed in constrained flows
in the Riga and Karlsruhe experiments. The case of experimental dynamos without internal
walls remains today an open question. The numerical study of von K\'arm\'an type flows 
shows the possibility to have self-generation of magnetic fields for magnetic 
Reynolds numbers accessible to a homogeneous sodium experiment. These flows 
appear to be very sensitive to the precise driving configuration and to boundary 
conditions. The comparison of the predictions concerning the induction effects  
with the data of the VKS experiment exhibits paradoxical results: the agreement is 
excellent in the case of one disk and intermediate $R_{\rm m}$ while the results
differ in the case of two counter-rotating disks. This effect could be 
due to the existence of the turbulent shear layer in the mid-plane, which is not 
accounted for in the mean velocity fields. 

\section*{Acknowledgements}
We would like to thank the VKS team (M. Bourgoin, J.B., A. Chiffaudel, F.D., S. Fauve, L.M., 
P. Odier, F. Petrelis, J.-F. Pinton) for fruitful discussions and 
to make available their not yet published data.
We are grateful to F. Stefani for providing us with numerical results
in a non-periodic cylindrical geometry.
J.B. thanks the Ministerio de Educaci\'on y Ciencia (spanish governement) for a 
post-doctoral grant when working at the CEA.

\vfill


\pagebreak

\begin{table}
\section*{Tables}
\end{table}

\begin{table}
\begin{tabular}{|c| c c |  c c| c c| c|}
\hline
                                           & 
\multicolumn{2}{c|}{ $U_{pol} (ms^{-1})$ } & 
\multicolumn{2}{c|}{ $U_{tor} (ms^{-1})$ } & 
\multicolumn{2}{c|}{ $U_{pol} / U_{tor}$ } &
$E_f$ \cr
& Mean & Max & Mean & Max & Mean & Max & \cr 
\hline
TM28 & 0.51 & 1.26 & 0.72 & 1.77 & 0.71 & 0.71 & 0.64\cr
TM60 & 0.47 & 1.20 & 0.58 & 0.94 & 0.82 & 1.27 & 0.52\cr
\hline
\end{tabular}
\caption{Characteristics of two experimental velocity fields corresponding to
propeller TM28 and TM60. The definitions
of mean and maximum velocity, poloidal and toroidal components
are given in section~\ref{velsection}. The efficiency corresponds to
$ E_f = U_{\rm max}/ 2\pi R_c f$.}
\label{ComparativeVitesse}
\end{table}


\onecolumn

\begin{figure}[htbp]
\section*{Figure Captions}
\end{figure}

\end{document}